\documentclass[12pt]{elsarticle}
\usepackage[top=0.8in, bottom=0.8in, left=0.8in, right=0.8in]{geometry}

\renewenvironment{thebibliography}[1]{
  \begin{oldthebibliography}{#1}
    \setlength{\itemsep}{0em}
    \setlength{\parskip}{0em}
}
{
  \end{oldthebibliography}
}

\usepackage[utf8]{inputenc} 
\usepackage[T1]{fontenc}
\usepackage{hyperref}
\usepackage{color}

\usepackage{amssymb}
\usepackage{amsmath}
\usepackage{changepage}

\usepackage{subcaption}
\usepackage{float}
\journal{Chaos, Solitons and Fractals}
\begin{document}

\begin{frontmatter}

\title{Community-level Contagion among Diverse Financial Assets}

\author[label1,label2]{An Pham Ngoc Nguyen\corref{*}}
\author[label1,label2]{Marija Bezbradica\corref{*}}
\author[label1,label2]{Martin Crane\corref{*}}
\affiliation[label1]{organization={ADAPT Center for Digital Content Technology},
            city={Dublin},
            postcode={D02 PN40},
            country={Ireland}}

\affiliation[label2]{organization={School of Computing, Dublin City University},
            addressline={Collins Ave Ext, Whitehall},
            city={Dublin},
            postcode={D09 V209},
            country={Ireland}}

\cortext[*]{Corresponding authors.\\Email addresses: ngocannguyen.pham@dcu.ie (A. P. N. Nguyen), marija.bezbradica@dcu.ie (M. Bezbradica),\\martin.crane@dcu.ie (M. Crane).}

\begin{abstract}
As global financial markets become increasingly interconnected, financial contagion has developed into a major influencer of asset price dynamics. Motivated by this context, our study explores financial contagion both within and between asset communities. We contribute to the literature by examining the contagion phenomenon at the community level rather than among individual assets. Our experiments rely on high-frequency data comprising cryptocurrencies, stocks and US ETFs over the 4-year period from April 2019 to May 2023. Using the Louvain community detection algorithm, Vector Autoregression contagion detection model and Tracy-Widom random matrix theory for noise removal from financial assets, we present three main findings. Firstly, while the magnitude of contagion remains relatively stable over time, contagion density (the percentage of asset pairs exhibiting contagion within a financial system) increases.  This suggests that market uncertainty is better characterized by the transmission of shocks more broadly than by the strength of any single spillover. Secondly, there is no significant difference between intra- and inter-community contagion, indicating that contagion is a system-wide phenomenon rather than being confined to specific asset groups. Lastly, certain communities themselves, especially those dominated by Information Technology assets, tend to act as major contagion transmitters in the financial network over the examined period, spreading shocks with high densities to many other communities. Our findings suggest that traditional risk management strategies such as portfolio diversification through investing in low-correlated assets or different types of investment vehicle might be insufficient due to widespread contagion.
\end{abstract}


\end{frontmatter}



\section{Introduction}\label{sec:1}
The financial markets have long been dominated by asset vehicles such as stocks, bonds, forex, ETFs and other traditional financial instruments, which serve as foundational components for investors and institutions alike \cite{mak24}. These markets are typically characterized by well-established regulatory frameworks, high liquidity and a vast body of research on asset dynamics and interdependencies \cite{ake10,che22,gon25}. In recent years, however, cryptocurrencies have emerged as a novel class of financial assets, gaining substantial attention due to their decentralized nature, high volatility, and increasing integration into mainstream financial systems~\cite{dro23}. Initially perceived as speculative instruments, cryptocurrencies such as Bitcoin and Ethereum are now showing signs of maturation, with growing market capitalization, institutional involvement and the development of derivative markets on the underlying cryptocurrency~\cite{aky20,aue23,jay24}. This evolving investment landscape presents new challenges and opportunities for understanding the behavior and interactions of various asset types within the financial ecosystem.

Investors participate in financial markets with the primary objective of generating returns on their investments. While the appeal of high profits often drives market engagement, it is undoubtedly accompanied by varying degrees of risk~\cite{cam05}. Risk is an inescapable and fundamental aspect of investing, stemming from uncertainties in market movements, macroeconomic factors, and asset-specific dynamics~\cite{cam05}. Even well-diversified portfolios cannot entirely eliminate exposure to risk~\cite{cam05}, making risk management a central concern in financial decision-making \cite{sha64}.

In light of the inherent risks associated with investing, a substantial body of literature has sought to uncover the underlying characteristics of financial markets to help investors make more informed decisions. These studies focus on enhancing investment strategies by identifying factors that can help in mitigating risks and optimizing returns~\cite{wka24,ngu25,ing25}. One significant area of this topic is the investigation of financial contagion - the phenomenon where economic shocks or market disturbances in one asset, market sector or country's financial market spread to other assets or sectors in that country or even other countries, leading to price fluctuations in the affected ones and potentially amplifying systemic risk~\cite{ell14}. Understanding these contagion mechanisms is crucial, as they can undermine diversification strategies and exacerbate market volatility. For example, a study conducted by Luchtenberg and Vu \cite{luc15} discovered the financial contagion effects between 10 countries from North America, Europe and East Asia Pacific during one of the biggest financial crashes in the Global Financial Crisis (GFC) of 2007/08. They revealed that this phenomenon became evident in all considered countries with key economic fundamentals such as trade structure, inflation, interest rates and investor risk aversion being significant determinants of contagion. The results suggest that international diversification by geography alone may not effectively shield investors, and that policymakers in both developed and developing countries must consider contagion risks in their financial strategies. In \cite{tan22}, the same pattern occurred when the contagion was shown to strike rapidly during the outbreak of the Covid-19 pandemic across global financial markets, especially in developing countries, eventually causing the synchronized downturns of numerous assets across various investment vehicles. To this end, our study contributes to this line of inquiry by examining the contagion dynamics across traditional assets such as stocks and ETFs as well as emerging ones like cryptocurrencies.

While much of the existing literature on financial contagion focuses on individual entities such as single securities, institutions or country-level financial markets~\cite{mat19,caf21,niy23,tor23}, there is a noticeable gap in understanding contagion dynamics at a broader structural level. Specifically, the phenomenon of contagion  within and between so-called 'communities' of similarly behaving assets (i.e. community-level contagion) remains underexplored. In the context of financial markets, a community refers to a cluster of assets that exhibit similar characteristics such as correlated price or volatility movements. Therefore, assets belonging to different communities exhibit more distinct characteristics compared to those within the same community~\cite{blo08}. These communities often emerge naturally in complex financial systems and can represent sectors, industries, functional groups of assets or financial events~\cite{dro20,wka242,sha21,ngu25}.  

Understanding contagion at the community level can yield valuable insights for investors and policymakers. First, it allows for improved risk assessment, as shocks may propagate more efficiently within highly interconnected communities and between densely linked communities, amplifying systemic risk~\cite{tor23}. Second, it enables more optimal diversification strategies which are unaffected by the contagion through the revelation of hidden interconnections between seemingly unrelated assets, especially when inter-community contagion is strong~\cite{lev70}. Lastly, identifying influential communities that initiate and spread shocks broadly to the rest of the network helps investors to monitor the financial markets more carefully and adjust portfolio allocations accordingly~\cite{ngu25}. By shifting the focus from individual to community-level contagion dynamics, this study aims to provide a more comprehensive understanding of financial contagion, thereby contributing to more robust investment and regulatory strategies. In particular, we seek to answer two research questions as follows:
\begin{enumerate}
    \item RQ1) Is the contagion within a community more pronounced than the contagion between different communities? Does the result suggest that contagion effects play a role in shaping community structures? Additionally, how does the result vary across different periods?
    \item RQ2) Is there evidence that some communities acting as central transmitters/receivers at certain times?
\end{enumerate}

This study draws on high-frequency data consisting of 221 time series of closing prices recorded at 30-minute intervals from April 2019 to May 2023. This period is chosen for its inclusion of several major events - both positive and negative - that significantly impact the financial markets. Moreover, it is also a time frame analyzed in related research \cite{ngu25}. The dataset covers 146 stocks, 49 U.S.-based exchange-traded funds (ETFs) tied to corresponding U.S. indices and 27 cryptocurrencies. All of this data was collected from the \href{https://firstratedata.com/}{FirstRate Data} platform. Regarding the methodologies, we use Minimum Spanning Tree (MST) and Louvain community detection to identify the community structure of the assets~\cite{ngu22,wka24}. The contagion signals are detected using the well-known Vector Autoregression (VAR) approach~\cite{lon10}. In addition,  measurement noise and inherent noise have been proven to exist in financial markets and potentially impact relevant experimental results~\cite{dim21,ngu22}. Therefore, to remove this effect from assets, we adopt the Tracy-Widom random matrix theory, which was introduced in \cite{tra94} and has been widely recognized as an efficient theory for noise removal in various areas, especially financial markets~\cite{din22,zhu22,apa20}. 

The remainder of the article is organized as follows: Section \ref{sec:2} presents an overview of related works. Section \ref{sec:3} provides a description of the datasets. Section \ref{sec:4} discusses methodology, preprocessing procedures and metrics. Section \ref{sec:5} describes the experimental results followed by their implications and hypothesis. Section \ref{sec:6} provides robustness analyses. Lastly, the conclusion of this study is given in Section \ref{sec:7}.

\section{Related Works} \label{sec:2}

\subsection{Upheaval in Financial Markets due to Recent Events}
\label{sec:2.1}

Global financial markets were affected by a series of major events during the 4-year period from April 2019 to May 2023. One of those that generated a great deal of headlines was the US-China trade war, which began in January 2018 with the US imposing tariffs on various Chinese imports, especially in the Information Technology sector~\cite{mol23}. In response, China enacted reciprocal tariffs on US goods and introduced legislation restricting the use of US-made technology products by Chinese consumers ~\cite{bow20,hou20}. The trade war lasted for more than 2 years until January 2020, peaking in intensity during 2019 and negatively affected financial markets in both countries. Sectors such as Commodities, Energy and Technology experienced sharp declines, with major stocks and ETFs suffering substantial losses \cite{apc21,che23}. Furthermore, as the central hubs of the global economy, the effects of this event spilled over to other parts of the world, increasing the volatility of financial markets worldwide \cite{ben23}.

The next noticeable event is the Covid-19 pandemic \cite{cio20}. Different studies have explored its impact across various aspects, such as asset correlations, investor behavior, policy responses and market reactions \cite{lah20,che21,mar21}. The same conclusion made by the majority is that the pandemic had an exceptionally strong and widespread effect on global financial markets, leading to one of the most significant economic crises in history. Specifically, this event caused a sharp price decline in a wide range of assets across nearly all types of investment vehicle \cite{lah20,che21}, with the most severe effects observed during the pandemic's first outbreak, between March and June 2020~\cite{lah20,che21}.  These consequences largely resulted from government interventions aimed at containing the virus, such as international border closures, mobility restrictions, and widespread shutdowns of economic activity \cite{mar21}.

Another event worth mentioning is the ongoing Ukraine-Russia conflict, which occurred on 24 February 2022 \cite{der24}. Russia, as a major exporter of oil and natural gas, while Ukraine, a key energy transit route and grain supplier, both play important roles in global supply chains, especially for countries in Europe. Consequently, the conflict between the two countries had a significant impact on financial markets, especially in relevant business sectors such as Energy, Financials and Consumer Staples \cite{hao23,jia24}. However, it is shown that the effects were uneven since the most severe financial consequences concentrated on European countries due to their strong commercial and political linkages with Russia and Ukraine\cite{gai22,ahm23}, while the impact on Asian markets and developed economies like the United States was limited or negligible \cite{bou222,you22,ass23}. Notably, the financial impact of this event was found to be modest compared to the past global disruptions like Covid-19 and the US–China trade war \cite{gai22,you22,wu23}.

Given such different influential events that occurred in financial markets during the period under consideration, we divide the initial dataset into sub-periods based on these events and market movements at that time. In fact, this division has been conducted by us in a previous study~\cite{ngu25}. This approach allows us to examine contagion effects in greater detail and under specific market conditions, thereby offering a more comprehensive understanding of the phenomenon.

\subsection{Contagion in Financial Markets}
\label{sec:2.2}

The topic of financial contagion has been extensively explored in the literature, with studies examining various dimensions of the phenomenon, such as contagion across countries, market indicators (e.g., interest rates, inflation rates, exchange rates, etc), trading exchanges and individual assets, both within the same investment type and across different types. For example, Ahmed and Masahiro in \cite{kha03} examined financial contagion among 9 Asian countries using 3 market indicators: stock market index, interest rate, and exchange rate against the US dollar. They employed the Vector Autoregressive (VAR) model to detect possible contagion effects during the period from July 1997 to June 1998, covering the 1997 Asian Crisis. Their findings revealed contagion in stock indices and currency exchange rates. However, the effect was weak and limited since not all examined countries showed signs of receiving or transmitting contagion while those that did only interacted with a few others. By contrast, they found no evidence of contagion in interest rates. Interestingly, the study noted that while contagion intensified during periods of high market turbulence, it was not the primary driver of the crisis in 1997 due to its weak and limited effect seen among the countries. 

As a common method for measuring contagion, Tugba et al. \cite{bas24} also used the VAR model for their research. In this study, they investigated the shock spillovers between cryptocurrencies, technology stocks, US ETFs and NFTs from March 2018 to September 2021. Their study assessed contagion both within and across different types of investment vehicle. The results showed that contagion primarily occurs within the same investment type since most contagion links are found among assets of the same type. However, these contagions do appear between technology stocks and US ETFs, while very little spillover is found between cryptocurrencies and traditional assets. In addition, NFTs show no contagion with other investment types. Focusing on a different aspect of contagion, Giudici and Pagnottoni in \cite{giu19} calculated the total spillover index (TSI) and directional spillover indices (DSI) using the Kwiatkowski–Phillips–Schmidt–Shin (KPPS) forecast error variance decompositions to examine hourly Bitcoin return spillovers among 5 cryptocurrency exchanges between July 2017 and June 2018. They emphasized that the spillovers among these exchanges constantly change over time. Remarkably, Bitfinex and Gemini exchanges are identified as leaders in price formation, transmitting return spillovers to the rest, while Bittrex tends to be a follower, receiving shocks from the others.

Apart from VAR, different methods have been introduced to examine financial contagion across different time periods and market conditions. For instance, Ahelegbey et al. introduced a new variant of the VAR model, known as Network VAR, in which the incorporation of interconnectedness networks into the original framework helps filter out unreliable contagion effects between time series \cite{ahe22}. Using this novel approach, they investigated the contagions among  20 major stock markets worldwide and further examined how these interconnections relate to global market risks over the period from 1999 to 2021. In a related study, the authors of \cite{niy23} employed DCC-GARCH (Dynamic Conditional Correlation GARCH) and Wavelet Cross-Correlation methods to examine the contagion across various cryptocurrencies and equity markets in both developing and giant economies from October 2014 to March 2022. Using a shorter timeframe from March 2017 to April 2020, Hao et al. \cite{wan22} looked at how contagion transmits back and forth between BTC, ETH and three US market indices - S\&P 500 (SPX), Dow Jones Industrial Average (DJI) and Nasdaq 100 (NDX) - using the Joe-Clayton copula GARCH model along with nonlinear Granger causality test. In another study \cite{mat19}, the regime switching skew-normal model was adopted to investigate the linear and non-linear contagion among big financial markets such as Euronext100, FTSE100 and Nikkei225 over a 3-year period from 2015 to 2018. Despite the differences in methodology and data, a consistent finding across these studies is that contagion intensifies during periods of heightened uncertainty, such as market crashes, major financial announcements and significant monetary policy changes.

Beyond analyzing contagion effects between individual assets, researchers have also turned their attention to understanding this phenomenon at the community level. For instance, \cite{tor23} investigated how different types of community structure formed from a set of banks influence the spread of liquidity contagion among the banks. The core research question tested is whether these community structures amplify or inhibit financial contagion. Using a micro-structural network model, the authors simulated how liquidity shocks propagate under various network topologies that differ in the strength of their community structure (e.g. the degree to which groups of nodes are tightly connected within their group and loosely connected to nodes outside their group). They found that the presence of communities generally increases both the number of banks affected and the overall liquidity shortfall, with contagion being most severe in networks with an intermediate level of community clustering. In this regard, the authors suggested that this result reveals a non-linear (U-shaped) relationship between community strength and contagion severity. In particular, a structure with individually strong and segregated communities can prevent shocks from spreading to the entire system, resulting in less overall contagion. However, as the community structure weakens to an intermediate level, shocks can reverberate within the remaining tighter communities, increasing their magnitude before spreading to the rest of the network, thus amplifying the initial shock. In contrast, shocks from a network with a weaker community structure dissipate rapidly without generating chain reactions. The study also tested the effectiveness of several policy interventions, including raising liquidity requirements across (i.e. among banks across different communities) or within specific communities (i.e. among banks within a community), assigning liquidity buffers based on centrality measures, and increasing market confidence through transparency. Results showed that higher liquidity requirements and improved market trust significantly reduce contagion, especially in tightly connected communities. Overall, this study emphasizes that regulators must consider the underlying community structures when assessing systemic risk and designing regulatory policies.

However, although this line of research has provided meaningful insights, it remains underexplored, especially in the context of widely traded financial assets such as stocks, ETFs, and cryptocurrencies. Most existing studies focus on contagion at the individual asset level or, if at the community level,  mainly within specialized domains like banking networks, leaving a gap in understanding how shocks propagate across interconnected communities of common financial instruments. To address this, our study aims to extend the contagion analysis to the community level across these asset types.

\subsection{Noise in Financial Assets and Noise Removal in Their Correlations} \label{sec:2.3}

A financial asset can be affected by measurement noise and inherent noise. Measurement noise arises during the data collection process where recorded prices deviate from true values due to factors such as non-synchronous trading, rounding and reporting inaccuracies~\cite{gri11}. In contrast, inherent noise is more complicated, stemming from the complex nature of assets and financial markets themselves~\cite{all06}. Indeed, this issue has been observed in both traditional and cryptocurrency markets, with the latter exhibiting a much higher degree of noise \cite{dim21}. One key contributor to this noise is the presence of noise traders~\cite{alf25} - typically less informed participants who attempt to imitate professional or popular traders, frequently relying heavily on headlines and market sentiments as well as lacking strong personal investment strategies \cite{ram15}. Consequently, their behavior can cause notable price volatility. Moreover, the dominance of na\"ive investors in the cryptocurrency market compared to more mature ones like equities or currencies also contributes to higher levels of noise~\cite{alm23}. Additionally, systemic risks such as new regulations and policy changes can also introduce further market disturbances. Studies have shown that both traditional assets and cryptocurrencies respond to such announcements within minutes and the resulting effects can persist for hours or even weeks \cite{aue18,yan23,ers21}. In the case of cryptocurrencies, noise might also come from phenomena such as \textit{Pump and Dump}, which is more prevalent in this market and serves as a distinguishing feature from traditional financial instruments \cite{li21}.


Noise in an asset has been shown to potentially hide its true underlying characteristics. In \cite{ait08}, NYSE stock prices were decomposed into fundamental components and microstructure noise using maximum likelihood estimation (MLE). The results indicated that even highly liquid assets contain noise, although the noise proportion tends to be less than low-liquidity ones. This implies that market prices of an asset may diverge from intrinsic prices based on the level of noise present. A similar finding was reported in the cryptocurrency market by Elie et al. \cite{bou24}, revealing that noise significantly affects prices, particularly in small-cap coins. In this regard, experiments using financial assets are more likely to be impacted by the noise. One notable example of this issue and also a key challenge faced in our present study is the distortion of asset correlations caused by noise. As shown in \cite{ngu22}, with the effect of noise, the correlations observed among a set of examined financial assets do not reflect their actual relationships. Moreover, the correlation patterns look very similar across different time periods and market circumstances. Once noise is filtered out, clearer correlation patterns emerge, differentiating stable from volatile periods -  a characteristic that was hidden by noise. 


Given the existence of noise in financial assets and its distortion in financial correlations, a widely used method for noise removal relies on Random Matrix Theory (RMT) and the Marchenko-Pastur distribution~\cite{ngu22,giu22}. This approach, introduced by Marchenko and Pastur, states that the eigenvalues of a purely random correlation matrix follow a specific distribution \cite{mar67}. Therefore, according to the theory, if all eigenvalues of an empirical matrix fall within this range, the matrix contains only noise. Conversely, eigenvalues outside this range are considered to carry meaningful information. Noise is removed by eliminating the eigenvalues within the Marchenko–Pastur bounds using techniques like Linear Shrinkage, Eigenvalue Clipping, or Rotationally Invariant Estimators \cite{bun17}. Indeed, this noise removal approach has been employed successfully in various studies. For instance, Laloux et al. in their portfolio optimization task denoised a correlation matrix of 406 S\&P500 stocks \cite{lal00} - a main variable in the Markowitz portfolio optimization algorithm, achieving better portfolio diversification with lower risk and higher returns compared to the original one. On the other hand, Laura et al. used it to identify safe-haven portfolios during market stress \cite{gon25}, while Hirdesh et al. found clearer correlation patterns, aiding in distinguishing different market states, such as stable and market crash \cite{pha19}. Furthermore, this method has also been used in other fields such as biology \cite{jal12}, medicine \cite{ade21}, and physics \cite{mit10}. However, a key limitation of this method is that it is sensitive to the length of the time period $T$ and the number of time series $N$. As a result, a change in either of them can affect the removal decision~\cite{paf03}. 


To address this limitation, an alternative approach using Tracy–Widom theory has been introduced~\cite{joh01}. In brief, Tracy and Widom in \cite{tra94} originally identified a fixed distribution for the largest eigenvalue of a random covariance matrix. Subsequent studies extended this to other eigenvalues due to their recurrent properties \cite{sac11}. As a result, each eigenvalue's distribution in a random matrix is characterized. Thus, given an empirical matrix, its noisy and informative eigenvalues can be distinguished by comparing them with corresponding random matrix-based theoretical distributions. Unlike Marchenko–Pastur, this approach is less sensitive to data size even in very small cases like $T = N = 5$ and especially shows robustness when $T \geq 200$ and $N \geq 200$ \cite{mil08}. Moreover, it incorporates empirical evidence, such as eigenvalue distributions obtained from thousands of simulations, into the noise removal process~\cite{sac11}. Therefore, it relies on observed distributions rather than strictly theoretical bounds as with the case of Marchenko–Pastur~\cite{sac11}, making it more robust and resilient. This theory has been adopted widely in different domains. For example, it was used for fault detection in the aviation field by comparing the largest empirical eigenvalue to the Tracy–Widom theoretical distribution, as shown in \cite{haj12}. In this regard, if the largest empirical eigenvalue falls within the 1\% critical region of the Tracy–Widom distribution, it is considered to deviate from typical noise behavior, suggesting faults in aircraft sensor systems. Similarly, in a biology-related study \cite{apa20}, the authors isolated meaningful biological signals by identifying and removing noisy eigenvalues from covariance matrices built from single-cell genomic data, revealing underlying characteristics hidden by the noise. In finance, several studies have demonstrated the effectiveness and robustness of Tracy-Widom theory in identifying noisy components within covariance or correlation matrices derived from financial asset data \cite{bou15,ake10}.

Leveraging the strengths of the Tracy–Widom-based random matrix theory, we employ this approach as a pre-processing step to filter out noise from the correlations among financial assets examined in our study. This denoising step is performed prior to all subsequent experiments. The detailed methodology for this process is provided in Subsections \ref{sec:4.2} and \ref{sec:4.3}.

\section{Dataset Overview} \label{sec:3}
\subsection{Time Series Description} \label{sec:3.1}
We utilize 221 time series sourced from the FirstRate Data platform, representing 146 stocks, 26 cryptocurrencies and 49 US ETFs. Each time series spans from 01/04/2019 to 03/05/2023. The original data was provided at tick-level granularity, requiring us to align a mutual resolution across all time series for our experiments. As a result, we upsampled all time series to 30-min intervals - the finest granularity that maintains their sufficient data availability. These 221 time series were selected based on the availability, liquidity and market capitalization, ensuring that only high-quality time series were included. Specifically, we apply the following criteria to select time series:
\begin{enumerate}
    \item Data availability: To ensure consistency in data sources, we focus on assets available on the FirstRate platform.
    \item Market capitalization: Given thousands of stocks available in the stock market, we restrict our selection to those whose underlying companies rank among the top 200 by market capitalization as listed on \href{https://companiesmarketcap.com/}{Companies Market Cap}. For US ETFs and cryptocurrencies, since each category has fewer than 100 assets available, we consider all of them. 
    \item High liquidity: We select assets with high trading frequency to ensure that historical price data is consistently available at a 30-min timescale, resulting in minimal missing values. Specifically, we include only stocks, US ETFs and cryptocurrencies with less than 1\%, 12\% and 10\% of missing values, respectively. These thresholds are chosen because assets exceeding them show significantly lower data availability. Nevertheless, the vast majority of assets have over 99\% actual data.  
\end{enumerate}

A list of assets in each category is shown in the supporting file S1.

\subsection{Timeline Division}\label{sec:3.2}

As noted in Section 2.1, the financial markets experienced several major disruptions during the four-year period from April 2019 to May 2023, including the US-China trade war, the Covid-19 pandemic and the political conflict between Ukraine and Russia. These different market conditions may exhibit varying contagion effects. To capture these potential dynamics, we divided the overall period into several segments aligned with the event timeline. Notably, since some segments are significantly longer than others, we further subdivided these into shorter intervals to prevent potential biases in the experimental results caused by the difference in time segment lengths, resulting in final segments of roughly equal length. It is our expectation that this segmentation facilitates a more balanced and finer-grained analysis. To this end, our experiments are conducted on each segment to compare how the contagion phenomenon varies across different market conditions.

Table \ref{tab:1} presents the seven segments (or sub-periods) used in this study. In addition to the three major events previously mentioned, the period after the Covid-19 outbreak and before the onset of the Ukraine-Russia conflict saw a significant surge in the prices of most assets \cite{wor22}. Accordingly, we refer to this interval as the Bull Time period.

\begin{table}[h!]
\centering
\caption{Timeline Division. The left column lists the name of each time segment while the right column indicates its corresponding duration.}
\resizebox{0.45\textwidth}{!}{%
\begin{tabular}{ll}

\hline
\textbf{Sub-period}                                                 & \textbf{Partition}       \\ \hline
Pre-Covid-19                                               & 01/04/2019 to 31/12/2019 \\ 
Covid-19 Outbreak                                                   & 01/01/2020 to 30/06/2020 \\ 
Bull Time 1                                                         & 01/07/2020 to 31/01/2021 \\ 
Bull Time 2                                                         & 01/02/2021 to 31/08/2021 \\ 
Bull Time 3                                                         & 01/09/2021 to 23/02/2022 \\ 
\begin{tabular}[l]{@{}l@{}}Ukraine-Russia\\ Conflict 1\end{tabular} & 24/02/2022 to 30/09/2022 \\ 
\begin{tabular}[l]{@{}l@{}}Ukraine-Russia\\ Conflict 2\end{tabular} & 01/10/2022 to 03/05/2023 \\ \hline
\end{tabular}}

\label{tab:1}

\end{table}

\subsection{Sector Division} \label{sec:3.3}
To gain a more detailed understanding of the assets and the contagion effects among them, we examine not only their types of investment vehicle but also their business sectors. Specifically, by convention, the stock market is classified into 11 categories based on business sectors, including  Communication Services, Utilities, Real Estate, Materials, Information Technology, Industrials, Healthcare, Financials, Energy, Consumer Staples, and Consumer Discretionary. Regarding US ETFs, these are more complex because each represents a basket of multiple stocks and reflects the overall performance of those underlying assets. As a result, a single US ETF may consist of stocks from various sectors. For instance, the well-known S\&P500 ETF (e.g. SPY) includes contributions from all sectors, with Technology being the largest one, accounting for 32\% of its holdings, as reported by \href{https://finance.yahoo.com/quote/SPY/holdings/}{Yahoo!finance}. For this characteristic, we assign a US ETF to a specific sector if it is designed to track the performance of stocks exclusively within that sector. Otherwise, it is classified as mixed. Cryptocurrencies, on the other hand, are not tied to specific business domains and are therefore treated as a single sector.  The distribution of stocks and US ETFs by business sector is shown in Table \ref{tab:2}.

\begin{table}[h!]
\centering
\caption{Classification of stocks (black text) and US ETFs (blue text) by business sector. Only US ETFs that track the performance of stocks exclusively within a sector are classified and considered to belong to that specific sector. Business sectors without available assets are not displayed.}
\resizebox{0.6\textwidth}{!}{%
\begin{tabular}{p{2.5cm} p{15cm}}
\hline
\textbf{Sector}                                                   & \textbf{Assets}                                                                                                                                                                                                                                                                        \\ \hline
Technology                                                        & ASX, AAPL, ACN, ADBE, ADI, ADP, AMAT, AMD, ASML, AVGO, BABA, CMCSA, CRM, CSCO, GOOG, IBM, INTC, INTU, LRCX, MA, META, MSFT, MU, NFLX, NOW, NVDA, ORCL, PANW, PDD, QCOM, SAP, SHOP, SONY, TMUS, TSM, TXN, T, VZ, V, \textcolor{blue}{CIBR},  \textcolor{blue}{IYW},   \textcolor{blue}{QTEC},  \textcolor{blue}{SOXX} \\ \\
Healthcare                                                        & ABBV, ABT, AMGN, AZN, BDX, BMY, BSX, CI, CVS, DHR, ELV, GILD, HCA, ISRG, JNJ, LLY, MDT, MRK, NVO, NVS, PFE, SNY, SYK, TMO, UNH, VRTX, ZTS,  \textcolor{blue}{IBB}                                                                                                                                         \\ \\
Financials                                                        & AXP, BAC, BLK, BX, CB, CFR, C, GS, HDB, HSBC, IBN, JPM, MC, MMC, MS, MUFG, SCHW, TD, WFC,  \textcolor{blue}{FTXO},  \textcolor{blue}{IYF},  \textcolor{blue}{KBWB}                                                                                                                                                                              \\ \\
Consumer Discretionary & ALV, AZMN, BKNG, COST, CSL, DIS, HD, IDEX, LOW, MCD, NKE, NVR, SBUX, TJX, TM, TSLA                                                                                                                                                                                                \\ \\
Consumer Staples     & BTI, BUD, DEO, EL, GE, HON, ITW, KO, MDLZ, MO, PEP, PG, PM, UL, WMT                                                                                                                                                                                                                    \\ \\
Industrials                                                       & ABB, AIR, BA, CAT, CNI, CP, DE, ETN, LMT, UNP, UPS, PYPL,   \textcolor{blue}{XTN}                                                                                                                                                                                                                \\ \\
Energy                                                            & BKR, BP, COP, CVX, EQNR, PBR, SHEL, SLB, TTE, EOG, XOM,  \textcolor{blue}{IEO},  \textcolor{blue}{IYE}                                                                                                                                                                                                                       \\ \\
Utilities                                                         & DTE, NEE, SO                                                                                                                                                                                                                                                                  \\  \\ 
Materials                                                         & BHP, LIN                                                                                                                                                                                                                                                                               \\ \\
Real Estate                                                       & AMT, PLD,  \textcolor{blue}{IYR}                                                                                                                                                                                                                                                                          \\ \hline
\end{tabular}}
\label{tab:2}
\end{table}

It is important to note that the number of stocks within each business sector varies and depends on the market capitalization of each asset. For example, sectors like Technology, Financials and Healthcare typically comprise stocks with high market capitalizations, whereas Communication Services and Real Estate are largely composed of lower-capitalization stocks \cite{dal24}. Consequently, sectors such as Utilities, Materials, Real Estate, and Communication Services contain a relatively limited number of entities. Given this limitation, our study focuses on the remaining seven sectors.

\section{Methodology} \label{sec:4}
\subsection{Asset Correlation Measurement} \label{sec:4.1}

The correlation between two assets is measured using the Pearson correlation coefficient~\cite{coh09}. This metric quantifies the linear relationship between two time series, with values ranging from -1 to 1. A value of 1 indicates a perfect positive correlation, meaning the two assets move in exactly the same way. By contrast, -1 indicates a perfect negative correlation, meaning the two assets move in opposite directions. A value of 0 suggests no linear relationship between the series. In essence, the higher the Pearson correlation, the more similar the movements of the two assets.

Given two log-return time series over a period of $T$ timestamps $x_i = (x_i^1, x_i^2, \cdots, x_i^T)$ and $x_j = (x_j^1, x_j^2, \cdots, x_j^T)$, corresponding to two assets $i$ and $j$, respectively, their Pearson correlation $c_{ij}$ is calculated as follows:
\begin{equation}
    c_{ij}  = \frac{\sum\limits_{t = 1}^T(x_i^t - \Bar{x_i})(x_j^t - \Bar{x_j})}{\sqrt{\sum\limits_{t=1}^T(x_i^t - \Bar{x_i})^2 \sum\limits_{t=1}^T (x_j^t - \Bar{x_j})^2}} \label{eq:1} 
\end{equation}

where log-return time series are transformed from original price time series \cite{tsa05}; $\Bar{x_i}$, $\Bar{x_j}$ are the average values of time series $x_i$, $x_j$, respectively. Subsequently, the correlation matrix $C$ for $N$ assets is constructed from all pairwise correlation values $c_{ij}$, capturing the relationship between every pair of assets. In other words, $C = (c_{ij})_{1 \leq i \leq N, 1 \leq j \leq N}$.

\subsection{Tracy-Widom Theory Recall} \label{sec:4.2}
As mentioned previously in Subsection \ref{sec:2.3}, we use the Tracy-Widom-based random matrix theory to remove noise from our empirical correlation matrices. Therefore, we first provide a brief overview of this theory to help readers better understand the approach before delving into the noise removal process in the next Subsection \ref{sec:4.3}.

Assume that the entries $x_{np}$ of a $n\times p$ random matrix $X$ are independent, identically distributed and follow the Gaussian distribution $N(0,1)$. Let $e_1$ be the largest eigenvalue of the corresponding $n\times n$ covariance matrix $Cov = XX^{\raisebox{0.5ex}{$\intercal$}}$. Declare two parameters
\begin{equation}
    u = (\sqrt{n-1} + \sqrt{p})^2 \label{eq:3}
\end{equation}

\begin{equation}
    s = (\sqrt{n-1} + \sqrt{p})\left(\frac{1}{\sqrt{n-1} + \frac{1}{\sqrt{p}}}\right)^{\frac{1}{3}} \label{eq:4}
\end{equation}

The standardized version $E_1$ of the largest eigenvalue $e_1$ is calculated as follows:
\begin{equation}
    E_1= \frac{e_1 - u}{s} \label{eq:4.2}
\end{equation}


With a set of $n \times n$ random covariance matrices constructed from the corresponding $n\times p$ random matrices, Tracy and Widom in \cite{tra02} have discovered that the standardized largest eigenvalues (as calculated in Equation \ref{eq:4.2}) of these covariance matrices converge to the following distribution function $F_1$ if $p/n \in (0,1]$ when $n \longrightarrow \infty$:
\begin{equation}
    F_1(s) = exp\left( -\frac{1}{2} \int_{s}^{\infty}q(t) + (t-s)q^2(t)dt\right) \label{eq:2}
\end{equation}

where $q(t)$ is the unique Hastings–McLeod solution \cite{has80} of the non-linear Painlevé differential equation $q^{''}(t) = tq(t) + 2q^3(t) $, satisfying the boundary condition $q(t) \approx Ai(t)$ when $t \rightarrow \infty$, with $Ai(t)$ being the Airy function \cite{air10}. A detailed description of this theory can be summarized as follows.

This method can be easily generalized to other eigenvalues thanks to a recurrence property as demonstrated in \cite{die05}. In other words, the distribution function for the standardized \textit{k}th largest eigenvalues from a set of random covariance matrices based on Tracy-Widom theory is known and one only needs to standardize these \textit{k}th eigenvalues by replacing the largest eigenvalue in Equation \ref{eq:4.2}. To this end, the Tracy-Widom theory provides a fixed distribution for each eigenvalue from a random covariance matrix.


However, we cannot adopt the presented theory since our study utilizes the correlation instead of the covariance matrix. Therefore, a transformation to adapt to the need of using correlation matrices has been discovered. For this, Edoardo et al. in \cite{sac11} have introduced another standardization procedure for the eigenvalues obtained from a correlation matrix so that the Tracy-Widom theory is still preserved in the case of correlation matrices. Without loss of generality, we introduce this standardization procedure for the largest eigenvalue. In particular, given the correlation matrix $Cov^*$ and its largest eigenvalue $e_1^*$, the distribution function $D$ of the largest eigenvalue $e_1^*$ can be experimentally constructed by generating a set of random correlation matrices that have similar characteristics as $Cov*$ (e.g. mean, standard deviation, size, etc). As a result, there exist parameters $a$ and $b$ such that the standardized eigenvalue $E_1^* = (e_1^* - a^*)/b^*$ follows the $F_1$ distribution. These parameters were theoretically derived by \cite{kol13} and have the following forms:
\begin{equation}
    a = m_{\tiny D}^1 - \frac{s_{\tiny D}^1}{s_{\tiny TW}^1}m_{\tiny TW}^1 \label{eq:5}
\end{equation}
\begin{equation}
    b = \frac{s_{\tiny D}^1}{s_{\tiny TW}^1} \label{eq:6}
\end{equation}

where $m^1_{{\tiny TW}},m_{\tiny D}^1$ and $s_{\tiny TW}^1,s_{\tiny D}^1$ are the mean and standard deviation of the Tracy-Widom distribution function $F_1$ and the experimental distribution $D$, respectively. Similar to the covariance matrices, this result can also be generalized to the remaining eigenvalues for which an appropriate distribution can be identified for each eigenvalue. In other words, the Tracy-Widom theory is preserved for all eigenvalues of a correlation matrix using this modified standardization.

To this end, this theory suggests that the eigenvalues of a random covariance or correlation matrix follow the Tracy-Widom distributions if standardized appropriately. Therefore, it can be used to identify whether an eigenvalue carries informative signals that reflect the underlying characteristics of the correlations between objects or simply carries random noise without real information. In this regard, an eigenvalue is considered to be informative if it falls outside the Tracy-Widom distribution. Otherwise, it is considered noise and will be removed~\cite{sac11}.

\subsection{Noise Filtering based on Tracy-Widom Theory} \label{sec:4.3}


Using the Tracy-Widom theory described in the previous section, we compare each eigenvalue of our empirical correlation matrix with the corresponding theoretical Tracy-Widom distribution of the eigenvalue to decide whether the eigenvalue is informative or simply noise. Specifically, given the eigenvalues $\lambda_i$ and eigenvectors $v_i$, with $1 \leq i \leq N$, obtained from a $N\times N$ correlation matrix $C$ such that $\lambda_1 > \lambda_2 > \cdots > \lambda_N$, we conduct a hypothesis test for each empirical eigenvalue $\lambda_i$ with the null hypothesis being the standardized eigenvalue $\lambda_i^*$ belongs to the theoretical Tracy-Widom distribution (i.e. the eigenvalue is noise). Moreover, we can say that, if the eigenvalue is too small compared to the Tracy-Widom distribution, it means that the information it carries is likely to be negligible even if the eigenvalue itself is located within the band where it is considered to contain information (e.g. in this case, the eigenvalue lies on the left tail of the Tracy-Widom distribution). Therefore, we considered these eigenvalues to be noise as well. In this regard, we set the significance level to 1\%, meaning that the null hypothesis is rejected (i.e. the eigenvalue is informative) if the standardized eigenvalue $\lambda_i^*$ falls within the rightmost 1\% tail of the Tracy-Widom distribution. Consequently, the informative eigenvalues of the correlation matrix $C$ are $\lambda_1, \cdots, \lambda_k$, where $k$ is the highest index such that the hypothesis test for $\lambda_{k+1}$ is accepted and the hypothesis tests for $\lambda_i, i \leq k$ are rejected.

Eventually, our denoising approach is to remove noisy eigenvalues from the correlation matrix $C$. For this, we employ the Eigenvector Clipping method~\cite{bou15} due to its key advantages. Firstly, it operates without the need for any training parameters, which enhances its robustness and reliability. In contrast, alternative denoising techniques such as Linear Shrinkage~\cite{bur22}, Non-linear Shrinkage~\cite{led12} and Rotationally Invariant Optimal Shrinkage~\cite{bun18} require various parameters' selection, raising the challenge of how to choose them appropriately. Secondly, this technique is easy to implement and efficient as it preserves the informative content of the data by keeping the trace of the correlation matrix unchanged after the cleaning process. Lastly, this method has demonstrated strong performance across multiple studies and has been successfully applied in diverse areas, including education, portfolio optimization, and signal processing~\cite{ngu23,con07,yan18}. The denoised correlation matrix $C_{denoised}$ based on the Eigenvector Clipping~\cite{bou15} is defined as follows:
\begin{equation}
    C_{denoised} = \sum_{i=1}^{n} \lambda_i^* \mathbf{v}_i \mathbf{v}_i^{\top}, \quad
\lambda_i^* = 
\begin{cases}
\dfrac{\lambda_{k+1} + \lambda_{k+2} + \cdots + \lambda_n}{n - k}, & \forall i \geq k + 1 \\
\lambda_i, & \forall i \leq k
\end{cases} \label{eq:8}
\end{equation}

\begin{figure}[h!]
    \centering
    \includegraphics[width=0.7\linewidth]{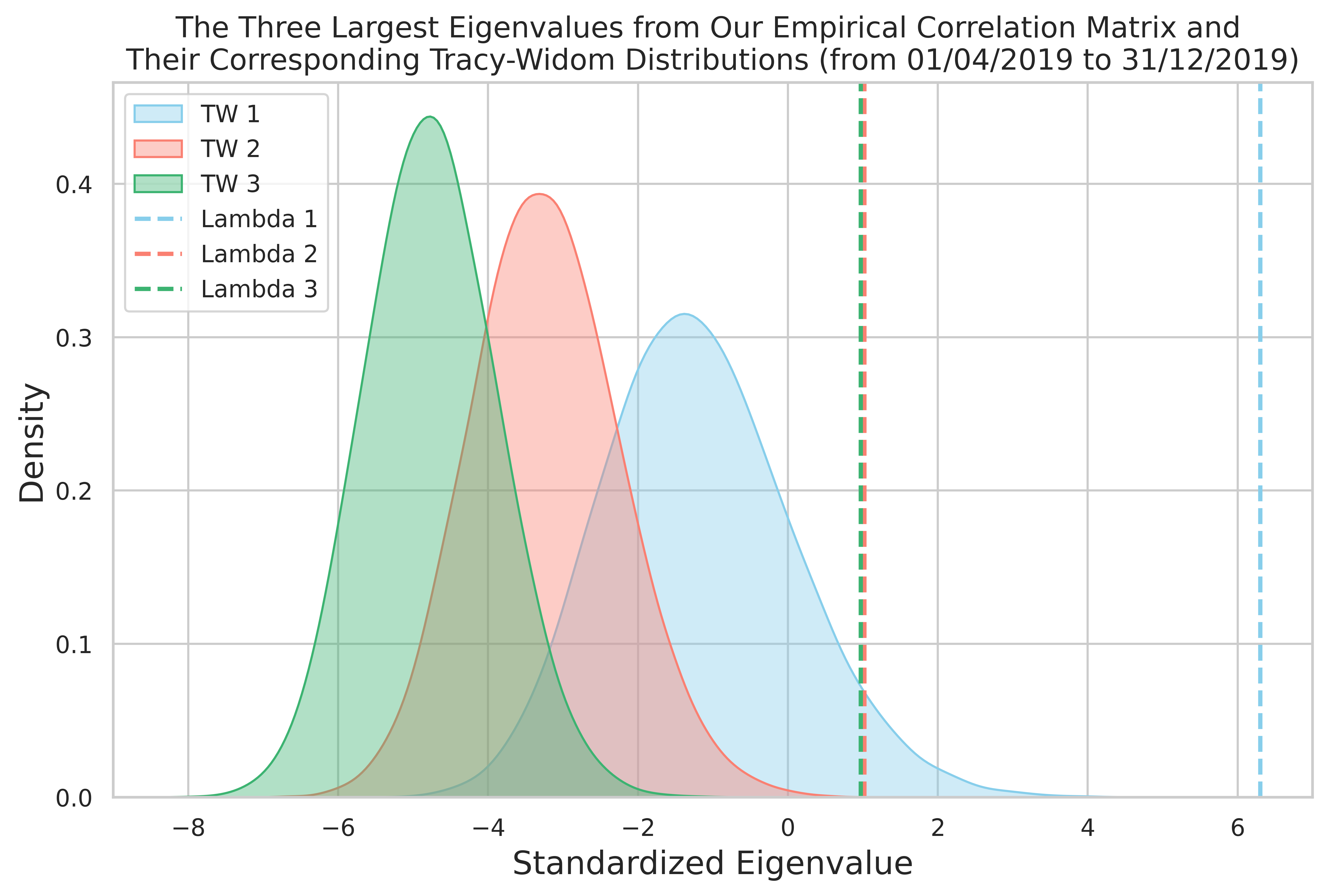}
    \caption{Three largest empirical eigenvalues (e.g. $\lambda_1 > \lambda_2 >\lambda_3$) and their corresponding Tracy-Widom distributions (e.g. TW 1, TW 2, TW 3) for the period from 01/04/2019 to 31/12/2019. All three eigenvalues are informative since they are outside and larger than the Tracy-Widom distributions. In other words, the three empirical eigenvalues lie within the rightmost 1\% tail (i.e. rejection area) of their corresponding Tracy–Widom distributions. This rejection area corresponds to the hypothesis that an eigenvalue is informative rather than noise.}
    \label{fig:1}
\end{figure}

\begin{figure}[h!]
    \centering
    \includegraphics[width=0.7\linewidth]{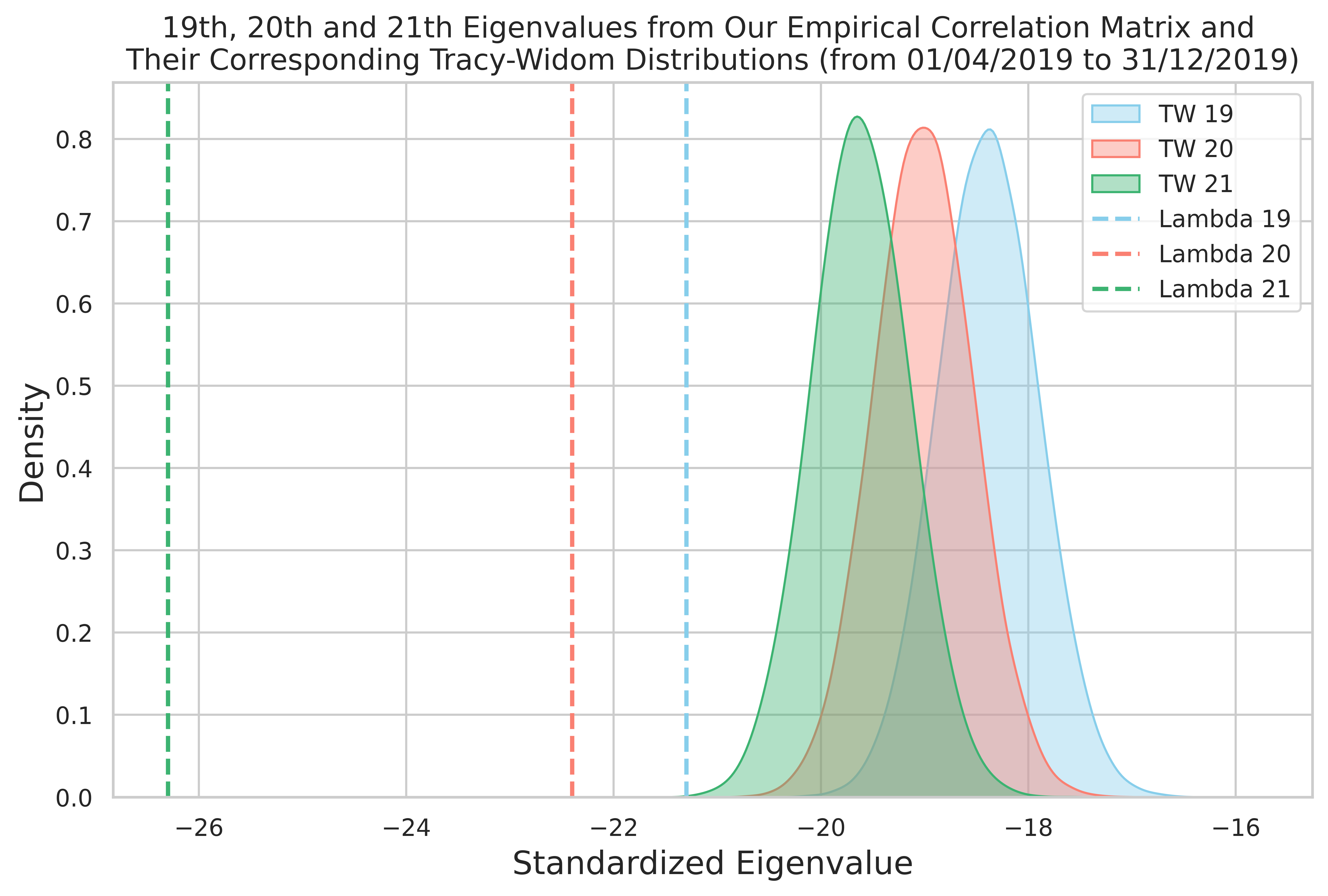}
    \caption{19th, 20th and 21st empirical eigenvalues (e.g. $\lambda_{19} > \lambda_{20} >\lambda_{21}$) and their corresponding Tracy-Widom distributions (e.g. TW 19, TW 20, TW 21) for the period from 01/04/2019 to 31/12/2019. Although these eigenvalues are outside the Tracy-Widom distribution, they are considered as noise since they are much smaller than the Tracy-Widom distributions, meaning that even if they carry real information, this amount is considered negligible. In other words, the three empirical eigenvalues lie outside the rightmost 1\% tail (i.e. rejection area) of their corresponding Tracy–Widom distributions.}
    \label{fig:2}
\end{figure}

An example of informative and noisy eigenvalues from our empirical correlation matrix identified by the Tracy-Widom theory is shown in Figures \ref{fig:1} and \ref{fig:2}, respectively. Each figure displays three examples in which the Tracy-Widom distribution for each eigenvalue is generated from 10,000 random correlation matrices and each vertical line represents a standardized eigenvalue from our empirical correlation matrix. Each empirical eigenvalue and its corresponding Tracy-Widom distribution have the same color. 

From our experiments in this present study, one common result is that the largest eigenvalues carry most of the information content of a correlation matrix, especially $\lambda_1$. For instance, this eigenvalue carries more than a third of the total information in the correlation matrix for the Pre-Covid period from 01/04/2019 to 31/12/2019. Notably, there are 221 eigenvalues in total, emphasizing the dominance of the largest eigenvalues. As a result, there is not much information left in the following eigenvalues, explaining the small magnitude of the 19th, 20th and 21st eigenvalues shown in Figure \ref{fig:2}. 

\subsection{Community Structure Construction} \label{sec:4.4}


From a correlation matrix, three preprocessing steps need to be conducted to get the community structure of the assets. Specifically, we first transform the correlation values $c_{ij} $ into the distance values $d_{ij} = \sqrt{2\times(1-c_{ij})}$. As a result, the correlation matrix $C$ is transformed into the so-called distance matrix $D_C = (d_{ij})_{1 \leq i \leq N, 1 \leq j \leq N}$. This transformation is necessary since the correlation values, although useful in understanding the relationships between entities, do not satisfy the requirements of a metric as defined in Algebra \cite{sea06}. Therefore, they cannot be used to make quantitative comparisons in terms of distance, which makes them unsuitable for directly evaluating how close or far entities are from each other in a measurable sense. Moreover, such a transformation is also mandatory for later algorithms that are only applicable in a metric space.   

An asset graph (or network) is then constructed from the distance matrix $D_C$ where each node of the graph (or network) represents an asset and an edge connecting two nodes represents their distance value. A major drawback of the graph lies in its high density where all nodes are connected with each other, making it both memory and computation-heavy. Since not all edges contribute equally useful information, we simplify the graph by removing edges with less important information. In particular, our second preprocessing step is using the Kruskal algorithm~\cite{kru56} to extract the Minimum Spanning Tree (MST) \cite{gra85} of the graph - a subgraph that preserves edges with the highest correlation degree (i.e. the smallest distance values) while maintaining the characteristics of a tree and discarding other edges. This allows us to retain the most important part of the graph while reducing the graph size and redundant information. Consequently, we reduce the graph from 221 nodes and 24,310 edges down to a much sparser structure with 221 nodes and just 220 edges - the ultimate network structure used in this study. 

Lastly, the third step involves applying the Louvain community detection algorithm~\cite{blo08} to the previously constructed MST in order to uncover the community structure among the assets. This algorithm operates in two phases based on modularity optimization, aiming to identify a partition that maximizes the similarity between nodes within each community while minimizing the similarity between the communities. This makes it particularly well-suited to our objective of grouping assets with similar price behavior and distinguishing them from less related ones.

\subsection{Contagion Detection} \label{sec:4.5}
Although different methods have been developed for contagion detection, the Vector Autoregression (VAR) model remains widely used and has consistently provided trustworthy insights in various studies \cite{kha03, bas24}. Therefore, we choose to use the VAR approach in our present study. Introduced by Sims \cite{sim80}, this is a statistical tool that helps analyze the relationships between several time series by expressing each variable as a function of its own past values and the past values of the other variables in the group \cite{lon10}. This allows the model to detect how shocks can move in both directions between any two time series. For example, a change in one time series may affect another and that second time series may in turn influence the first. Given a set of $m$ time series $\{y_1, y_2, \cdots, y_m\}$, the VAR model of order $p$ can be written as:

\begin{align}
    \begin{bmatrix}
y_1^t \\
y_2^t \\
\vdots \\
y_m^t
\end{bmatrix} =  \begin{bmatrix}
c_1 \\
c_2 \\
\vdots \\
c_m
\end{bmatrix} &+ \begin{bmatrix}
a_{11}^{t-1} & \cdots & a_{1m}^{t-1} \\
a_{21}^{t-1} & \cdots & a_{2m}^{t-1} \\
 & \vdots & \\
a_{m1}^{t-1} & \cdots & a_{mm}^{t-1}
\end{bmatrix}
\begin{bmatrix}
y_1^{t-1} \\
y_2^{t-1} \\
\vdots \\
y_m^{t-1}
\end{bmatrix} \nonumber \\ &+ \dots + \begin{bmatrix}
a_{11}^{t-p} & \cdots & a_{1m}^{t-p} \\
a_{21}^{t-p} & \cdots & a_{2m}^{t-p} \\
 & \vdots & \\
a_{m1}^{t-p} & \cdots & a_{mm}^{t-p}
\end{bmatrix}
\begin{bmatrix}
y_1^{t-p} \\
y_2^{t-p} \\
\vdots \\
y_m^{t-p}
\end{bmatrix}
+
\begin{bmatrix}
\varepsilon_1 \\
\varepsilon_2 \\
\vdots \\
\varepsilon_m
\end{bmatrix}\label{eq:9}
\end{align}

where $y_i^t$ represents the value of time series $y_i$ at time $t$. The letter $p$ indicates how many previous time steps (lags) are considered when predicting the current value. Each time series depends not only on its own past values (up to $p$ lags), but also on the past values of all other time series in the system over the same number of lags. The term $c_i$ is a constant (intercept) in the equation for the time series $y_i$. The coefficient $a_{i,j}^{t-z}$ represents the contagion effect of time series $y_j$ on $y_i$ at lag $z$. If this coefficient is statistically significant at the 90\% confidence level or higher, we consider that there is a contagion effect from $y_j$ to $y_i$ at lag $z$. Furthermore, a positive value means that an increase in one time series tends to be followed by an increase in the other, while a negative value suggests the opposite effect. Finally, $\epsilon_i$ is the error term in the equation for $y_i$, capturing random fluctuations not explained by the model.

We acknowledge that using a stricter confidence level such as 95\% or 99\% can provide more rigorous analytical results.  However, our goal is to keep as many contagion coefficients as possible to ensure a sufficient sample size within and between communities for statistical testing. Adopting such stricter thresholds would reduce the number of available contagion coefficients, potentially leaving some tests with too few observations and thereby undermining their statistical validity. In contrast, the 90\% confidence level remains a widely accepted standard in the literature and allows us to retain a higher number of observations. To this end, we chose to use the 90\% confidence level for this analysis.

From our experiments, we found that the contagion effects between assets differ across lags, with each lag offering a different perspective. In this study, we focus specifically on short-term contagion, represented by 30-min lag effects. Future research will extend this work to include longer time horizons.

\subsection{Contagion Metrics} \label{sec:4.6}
We evaluate community-level contagion effects in two ways: overall and directional. The overall effect captures how shocks spread in both incoming and outgoing directions (e.g. bi-directional) among the assets within a community or between two communities. On the other hand, the directional effect measures how contagion flows from one community of assets to another (e.g. unidirectional). Based on these perspectives, we define six metrics outlined below. The design of these metrics is inspired by \cite{asi21} and \cite{mat13}.
\begin{enumerate}
    \item Overall Contagion Density within a Community($OCD_{individual}$): to measure the density of contagion within a community of assets, considering both incoming and outgoing directions. Let $S$ be a set of contagion links between $n$ assets.
    \begin{equation}
        OCD_{individual} = 100 * \frac{L_S}{n \times (n-1)} \label{eq:10}
    \end{equation}
    where $L_S$ is the cardinality of $S$ and $n\times (n-1)$ is the number of possible contagion links between $n$ assets
    \item Overall Contagion Magnitude within a Community ($OCM_{individual}$): to measure the average magnitude of contagion within a community of assets, reflecting how strong the contagion is, on average, within the community. Similar to the $OCD_{individual}$ metric, this considers both incoming and outgoing directions.
    \begin{equation}
        OCM_{individual} = \frac{\sum_{(i,j) \in S} w_{ij}}{L_S} \label{eq:11}
    \end{equation}
    where $\sum_{(i,j) \in S} w_{ij}$ is the sum of the contagion magnitude in $S$. 
    \item Overall Contagion Density between Two Communities ($OCD_{cross}$): to measure the density of contagion between communities, considering both incoming and outgoing directions. Let $SA$ and $SB$ be the sets of contagion links from $n$ assets in community $A$ to $m$ assets in community $B$ and vice versa, respectively.
    \begin{equation}
        OCD_{cross} = \frac{L_{SA} + L_{SB} }{2\times m \times n} \label{eq:12}
    \end{equation}
    where $L_{SA}$ and $L_{SB}$ are the cardinality of $SA$ and $SB$, respectively. $2\times m \times n$ is the number of possible contagion links between two communities.
    \item Overall Contagion Magnitude between Two Communities ($OCM_{cross}$): to measure the average magnitude of contagion between two communities, considering both incoming and outgoing directions.
    \begin{equation}
        OCM_{cross} = \frac{\sum_{(i,j) \in SA} w_{ij} + \sum_{(l,z) \in SB} w_{lz}}{L_{SA} + L_{SB}} \label{eq:13}
    \end{equation}
    where $\sum_{(i,j) \in SA} w_{ij}$ and $\sum_{(l,z) \in SB} w_{lz}$ are the sum of the contagion magnitude in $SA$ and $SB$, respectively.
    \item Directional Contagion Density (DCD): to measure the density of contagion from a community A to a community B. Let $A$ and $B$ be two communities with $n_A$ and $n_B$ assets, respectively.
    \begin{equation}
        DCD = 100 * \frac{L_{A \rightarrow B}}{n_A \times n_B} \label{eq:14}
    \end{equation}
    where $L_{A \rightarrow B}$ is the number of contagion links directed from assets in $A$ to assets in $B$ and $n_A\times n_B$ is the number of possible directed contagion links from $A$ to $B$.
    \item Directional Contagion Magnitude (DCM): to measure the average magnitude of contagion from a community $A$ to a community $B$, reflecting how strong the directed contagion is, on average, from $A$ to $B$
    \begin{equation}
        DCM = \frac{\sum_{(i\in A, j\in B)} w_{ij}}{L_{A \rightarrow B}} \label{eq:15}
    \end{equation}
    where $w_{ij}$ is the magnitude of contagion from asset $i$ in $A$ to asset $j$ in $B$.
\end{enumerate}

\section{Results and Implications} \label{sec:5}
This section presents and discusses the experimental results on contagion effects during each of the seven aforementioned sub-periods, spanning from 01/04/2019 until 03/05/2023. We recall that this study focuses on contagion at the community level, investigating the phenomenon between communities of assets rather than between individual assets. These communities are identified using the Louvain community detection algorithm, with the Tracy-Widom theory applied to mitigate the impact of noise. Further details of this process are provided in Section \ref{sec:4}. The structure of this section follows the two defined research questions and is divided into 4 subsections. First, we provide an overview of the network and community structure constructed from 221 assets across seven sub-periods. Next, we explore the contagion effects both within individual communities and among each pair of communities. This experiment aims to determine whether intra-community contagion differs significantly from inter-community contagion, answering the first research question. Then, we examine whether certain communities act as dominant transmitters (i.e. transmitting shocks to the rest of the network) or receivers (i.e. receiving shocks from the rest of the network) of contagion signals, thereby addressing the second research question. Finally, implications gathered from the results will be provided in the fourth subsection.

Although our study covers several major financial and global events, it does not capture the most recent structural changes in traditional and cryptocurrency markets. Given the rapid evolution of market microstructure, regulatory frameworks and financial conditions since 2023, extending the dataset to include more recent observations could provide additional insights and further validate our results. While incorporating such data lies beyond the scope of the present study, we acknowledge this as a limitation and encourage future research to integrate up-to-date information to enhance and extend our findings.

\subsection{Network and Community Structures Discovery} \label{sec:5.1}
\begin{figure}[h!]
\centering
\begin{subfigure}{\linewidth}
\centering
\includegraphics[width= 11cm]{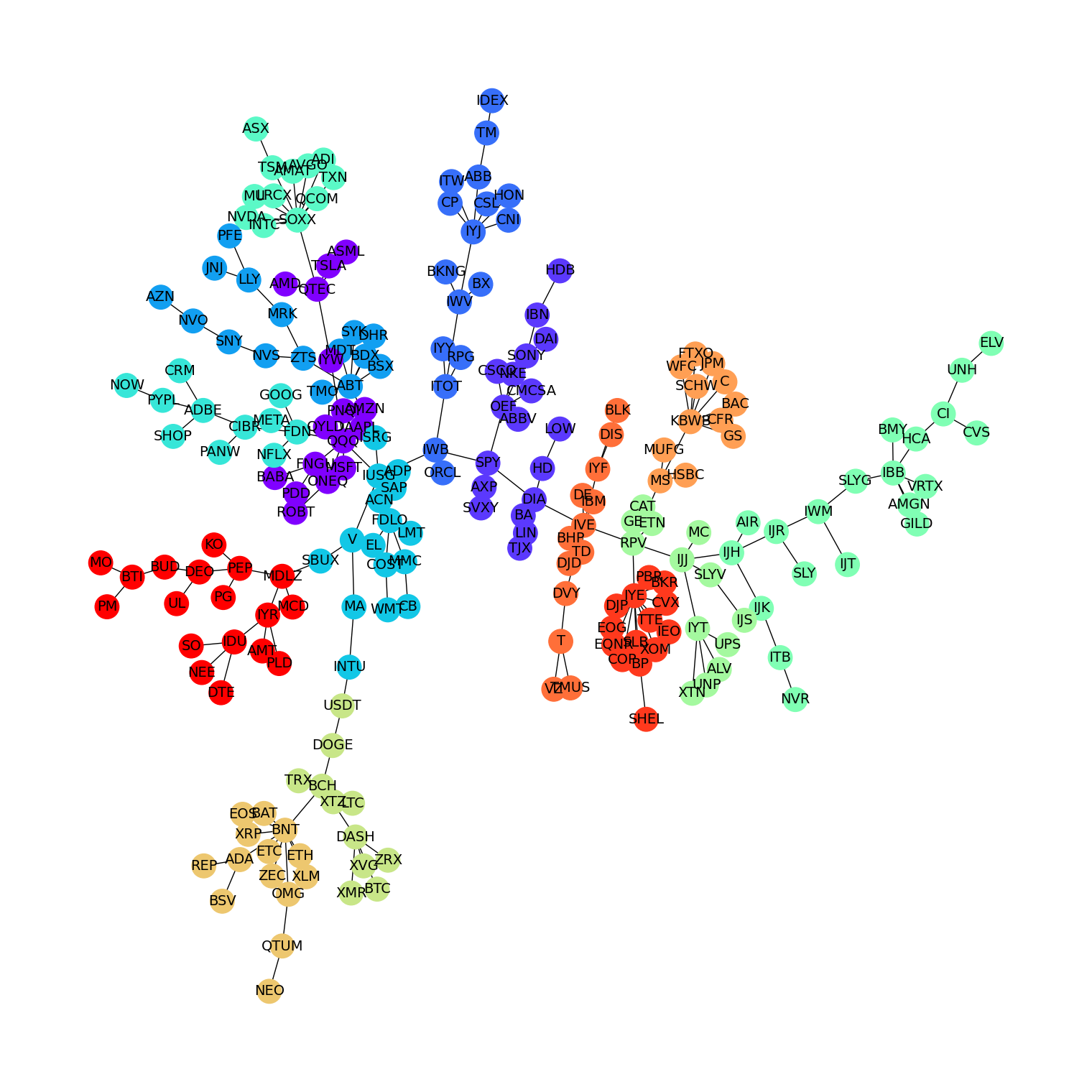}
\caption{Network and Community Structure}\label{fig3.1}
\end{subfigure}%

\begin{subfigure}{\linewidth}
\centering
\includegraphics[width= 11cm]{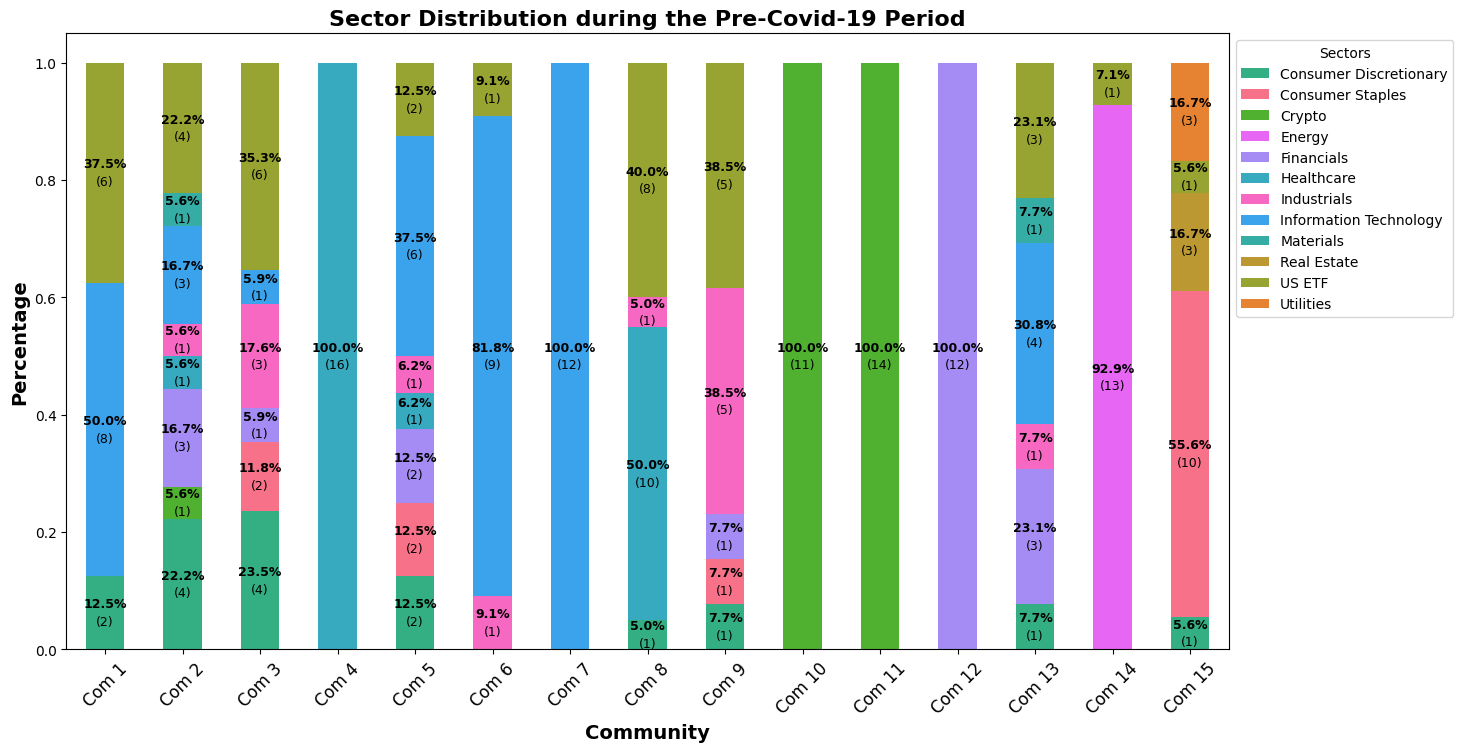}
\caption{Sector Distribution}\label{fig3.2}
\end{subfigure}

\caption{Network and Community structure (a) and sector distribution per community (b) of 221 assets during the Pre-Covid-19 period. A US ETF is assigned to a specific business sector if it is designed to track the performance of stocks within that sector. Otherwise, it is unclassified and labeled as US ETF (i.e. US ETFs that do not track a specific sector or those that include stocks from multiple sectors). \label{fig:3}}
\end{figure}

\begin{figure}[h!]
\centering
\begin{subfigure}{\linewidth}
\centering
\includegraphics[width= 11cm]{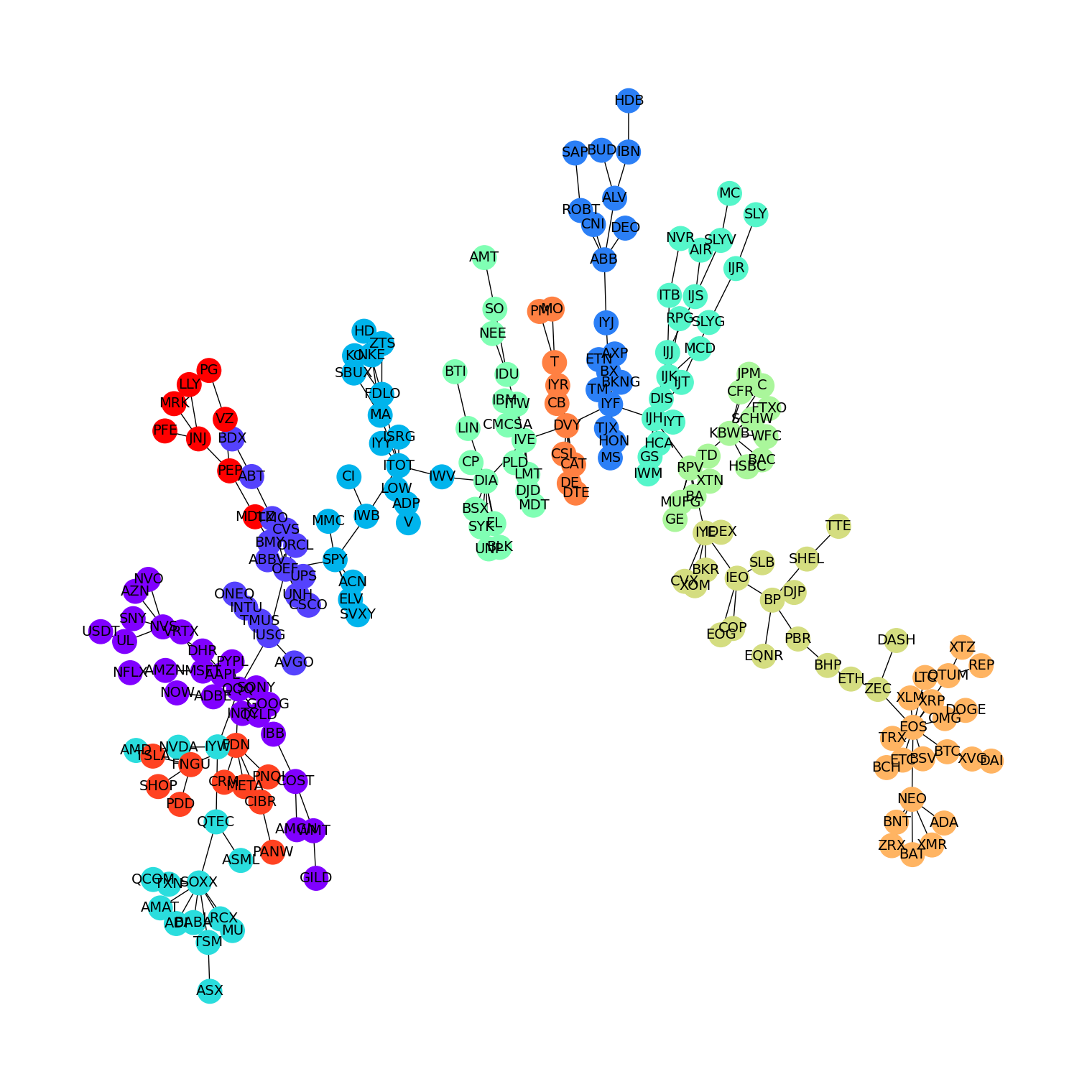}
\caption{Network and Community Structure}\label{fig4.1}
\end{subfigure}%

\begin{subfigure}{\linewidth}
\centering
\includegraphics[width= 11cm]{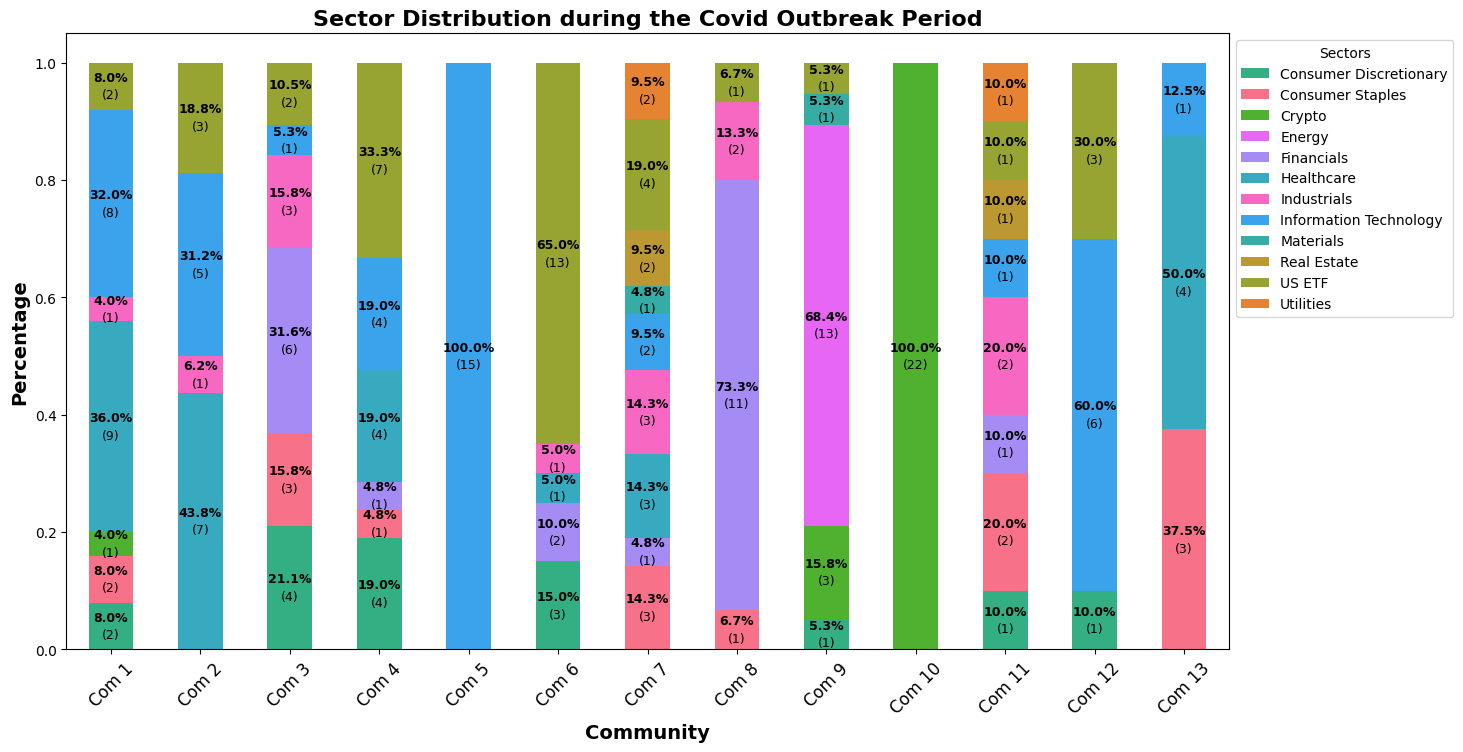}
\caption{Sector Distribution}\label{fig4.2}
\end{subfigure}

\caption{Network and Community structure (a) and sector distribution per community (b) of 221 assets during the Covid-19 Outbreak period. A US ETF is assigned to a specific business sector if it is designed to track the performance of stocks within that sector. Otherwise, it is unclassified and labeled as US ETF (i.e. US ETFs that do not track a specific sector or those that include stocks from multiple sectors) \label{fig:4}}
\end{figure}

We first examine the network and community structure of 221 assets drawn from 3 categories: stocks, cryptocurrencies and US ETFs, across seven sub-periods, namely Pre-Covid-19, Covid-19 Outbreak, Bull Time 1, Bull Time 2, Bull Time 3, Ukraine-Russia Conflict 1 and Ukraine-Russia Conflict 2. These network and community structures will be used for experiments in the subsequent subsections. Examples of the network as well as community structures are displayed in Figures \ref{fig3.1} and \ref{fig4.1} while full details for all sub-periods are provided in the supporting file S2. In these graphs, each node represents an asset and an edge between two nodes indicates a significant similarity (i.e. strong correlation) between them. On the other hand, the absence of an edge suggests a weak or negligible similarity relative to the surrounding connections. In a network, each community is highlighted by a different color. A detailed breakdown of each community within each sub-period is provided in the supplementary file S3.

To quantitatively assess the network structure of assets in each sub-period, we employ three widely used network centrality measures, including Degree Assortativity \cite{nol15}, Betweenness Centrality \cite{bor05}, and Closeness Centrality \cite{bor05}. In terms of community structures, we also use Betweenness Centrality and Closeness Centrality but on each individual community (i.e. Community Betweenness and Community Closeness) rather than the whole network like the former. We note that the results of each metric will be averaged to get an overall value for each sub-period. While detailed definitions of these measures can be found in the cited references, a brief explanation is as follows: lower values of Degree Assortativity and Betweenness Centrality indicate a more compressed network, where assets are relatively close to one another and tend to form large communities. Conversely, higher values suggest a more distributed structure, with assets further apart and tend to form a larger number of smaller communities. In contrast, the interpretation of Closeness Centrality is reversed: higher values correspond to a more compressed network.

The results of this assessment are presented in Table \ref{tab:revise_1}. Overall, the values of each metric on the network structure vary across sub-periods and no single sub-period consistently exhibits the most compressed or most distributed structure. This indicates that the network structure evolves over time and seems not to be influenced by market conditions. At the community level, although the network centrality measures also fluctuate over time, the Covid-19 Outbreak sub-period consistently shows a more compressed structure compared with the others. This indicates that assets became more closely connected within their communities during this time, implying that information flowed more rapidly and efficiently among assets as a result of the market turbulence triggered by the Covid-19 pandemic.

\begin{table}[h!]
\caption{Network centrality measures in each sub-period}
\label{tab:revise_1}
\centering
\resizebox{1.0\textwidth}{!}{
\begin{tabular}{cccccccc}
\hline
\textbf{Network Centrality Measure} & \textbf{Pre-Covid-19} & \textbf{\begin{tabular}[c]{@{}c@{}}Covid-19\\ Outbreak\end{tabular}} & \textbf{Bull Time 1} & \textbf{Bull Time 2} & \textbf{Bull Time 3} & \textbf{\begin{tabular}[c]{@{}c@{}}U-R\\ Conflict 1\end{tabular}} & \textbf{\begin{tabular}[c]{@{}c@{}}U-R\\ Conflict 2\end{tabular}} \\ \hline
Degree Assortativity & -0.435 & -0.422 & -0.354 & -0.368 & -0.355 & -0.391 & -0.397 \\
Betweenness Centrality & 0.038 & 0.043 & 0.042 & 0.050 & 0.048 & 0.041 & 0.040 \\
Closeness Centrality & 0.127 & 0.138 & 0.123 & 0.101 & 0.112 & 0.139 & 0.135 \\ 
Community Betweenness & 0.139 & 0.125 & 0.157 & 0.160 & 0.167 & 0.144 & 0.150 \\
Community Closeness & 0.429 & 0.516 & 0.438 & 0.423 & 0.457 & 0.492 & 0.465 \\ \hline
\end{tabular}}
\end{table}

However, although the network and community structure changes across sub-periods with assets clustering and dispersing in different ways on a sub-period basis, some patterns remain unchanged over time. In particular, it can be seen that cryptocurrencies tend to separate themselves from traditional assets, distributing close to each other and forming distinct communities. Exceptions are USDT and DAI - two popular stablecoins designed to track the USD currency, which explains their closer association with traditional markets compared to the cryptocurrency one \cite{wu232}. By contrast, stocks and US ETFs are more likely to cluster together, emphasizing a strong relationship between them. To support this observation, we measure the average distance between assets within each category and between assets from two different categories, then compare them with each other. Specifically, using the Minimum Spanning Tree (MST) and its distance values computed in subsection \ref{sec:4.4}, we calculate the pairwise distance between each pair of assets as the sum of the edge weights along the shortest path connecting them. We then compute the average intra-category distances (e.g. stocks versus stocks) and inter-category distances (e.g. stocks versus cryptocurrencies). This results in 6 pairs of comparison, including cryptocurrencies with themselves, stocks with themselves, US ETFs with themselves, cryptocurrencies versus stocks, cryptocurrencies versus US ETFs and stocks versus US ETFs. 

As shown in Table \ref{tab:3}, the average distance among cryptocurrencies is significantly lower than in other cases, revealing their strong internal cohesion. Similarly, stocks and US ETFs also show moderate distances among each other, reflecting a meaningful degree of similarity. In contrast, the average distance between a cryptocurrency and a traditional asset is approximately 4 times higher than the crypto-crypto average distance and about twice that of the stock-ETF average distance. This result reinforces the finding that stocks and US ETFs are close to each other and tend to mix within the same communities while cryptocurrencies show a distinction to some extent, forming their own communities and displaying less similarity to traditional assets.
\begin{table}[h!]
\centering
\caption{Average distance between assets within a category and between assets from two different categories}
\label{tab:3}
\resizebox{0.7\textwidth}{!}{%
\begin{tabular}{ccccccc}
\hline
\textbf{Period} & \multicolumn{1}{c}{\textbf{\begin{tabular}[c]{@{}c@{}}Crypto\\ vs\\ Crypto\end{tabular}}} & \multicolumn{1}{c}{\textbf{\begin{tabular}[c]{@{}c@{}}Stock\\ vs\\ Stock\end{tabular}}} & \multicolumn{1}{c}{\textbf{\begin{tabular}[c]{@{}c@{}}US ETF\\ vs \\ US ETF\end{tabular}}} & \multicolumn{1}{c}{\textbf{\begin{tabular}[c]{@{}c@{}}Crypto\\ vs\\ US ETF\end{tabular}}} & \multicolumn{1}{c}{\textbf{\begin{tabular}[c]{@{}c@{}}Crypto\\ vs\\ Stock\end{tabular}}} & \multicolumn{1}{c}{\textbf{\begin{tabular}[c]{@{}c@{}}Stock\\ vs\\ US ETF\end{tabular}}} \\ \hline
Pre-Covid-19 & 3.605 & 7.944 & 5.278 & 11.556 & 12.150 & 6.763 \\ 
Covid-19 Outbreak & 3.583 & 6.694 & 5.259 & 11.423 & 12.807 & 6.058 \\ 
Bull Time 1 & 4.675 & 7.863 & 6.137 & 12.040 & 13.131 & 7.121 \\ 
Bull Time 2 & 3.617 & 10.276 & 6.795 & 12.641 & 13.750 & 8.665 \\ 
Bull Time 3 & 3.362 & 8.383 & 5.625 & 16.502 & 16.015 & 7.290 \\ 
Ukraine-Russia Conflict 1 & 3.224 & 7.766 & 4.940 & 8.713 & 9.330 & 6.595 \\ 
Ukraine-Russia Conflict 2 & 3.406 & 7.575 & 4.335 & 10.761 & 12.517 & 6.034 \\ \hline
\end{tabular}}
\end{table}

Remarkably, stocks and US ETFs from certain business sectors, including Information Technology, Healthcare, Financials and Energy, tend to form distinct communities among themselves or at least account for the majority within a community. To reinforce this result, we analyze the business sector distribution of each community, where the percentage of each sector within an individual community is calculated. Representative results are shown in Figures \ref{fig3.2} and \ref{fig4.2}. Specifically, during the Pre-Covid-19 sub-period (Figure \ref{fig3.2}), apart from 2 communities composed fully of cryptocurrencies (e.g. communities \#10 and \#11), stocks and US ETFs in the Information Technology sector form the community \# 7 and are dominant in communities \#1 and \#6. Likewise, assets from the Healthcare, Financials and Energy sectors also show similar behavior, forming communities \#4, \#12 and \#14, respectively. Likewise, in the next sub-period (Figure \ref{fig4.2}), assets in Information Technology, Financials and Energy account for the majority of communities \#5, \#8 and \#9, respectively. This sector-based clustering pattern is consistently observed across all sub-periods. From these results, we seek to answer the following key questions: 1) How do intra-community contagion effects compare to inter-community contagion effects? 2) Does this comparison provide evidence that strong contagion among a set of assets drives community formation between them, particularly for assets within the same sector? 3) Are there specific communities that serve as dominant transmitters or receivers of contagion across the network? These questions will be addressed in the next subsections. By understanding contagion effects in a broad market that includes various types of assets, investors can better control their investment portfolios to minimize potential risks and take advantage of market movements. For instance, avoiding investments in assets with dense contagion can help reduce systemic risk in a portfolio. Additionally, identifying the sources of contagion can help investors anticipate potential changes in portfolio performance, enabling them to make timely adjustments and take advantage of informed insights.

\subsection{Intra-community vs Inter-community Contagion} \label{sec:5.2}
In this subsection, we evaluate the community-level contagion in two ways: within a single community (intra-community contagion) and between two different communities (inter-community contagion). To do this, we apply the overall contagion metrics defined in Section \ref{sec:4.6}, which account for both incoming and outgoing contagion directions, as described in Equations \ref{eq:10}-\ref{eq:13}. The main objective is to determine whether contagion effects are stronger among assets within the same community compared to those between different communities, which is equivalent to the first research question. Additionally, from experiments in this section, we also investigate several related questions: are there any communities that show significantly strong intra-contagion in the network? Similarly, are there any pairs of communities that exhibit significantly strong inter-contagion between them? What factors might drive this behavior? How stable are the contagion results across different time periods and market conditions? 

Except for the Covid-19 Outbreak sub-period with 13 communities forming from 221 assets, other sub-periods have at least 15 communities, leading to more than 100 inter-community contagions in each sub-period. Given this huge amount of information, we have to come up with a way to interpret it all. With this in mind, for each sub-period, we construct a distribution of inter-community contagions, thereby the contagion effects between pairs of communities can be understood transparently in each sub-period while being beneficial for further experiments conducted in this study. Additionally, we also construct a distribution for intra-community contagions in each sub-period to facilitate the comparison of intra-community contagion effects across sub-periods, which will be presented subsequently.

\begin{figure}[h!]
\centering
\begin{subfigure}{0.5\linewidth}
\centering
\includegraphics[width= 8.5cm]{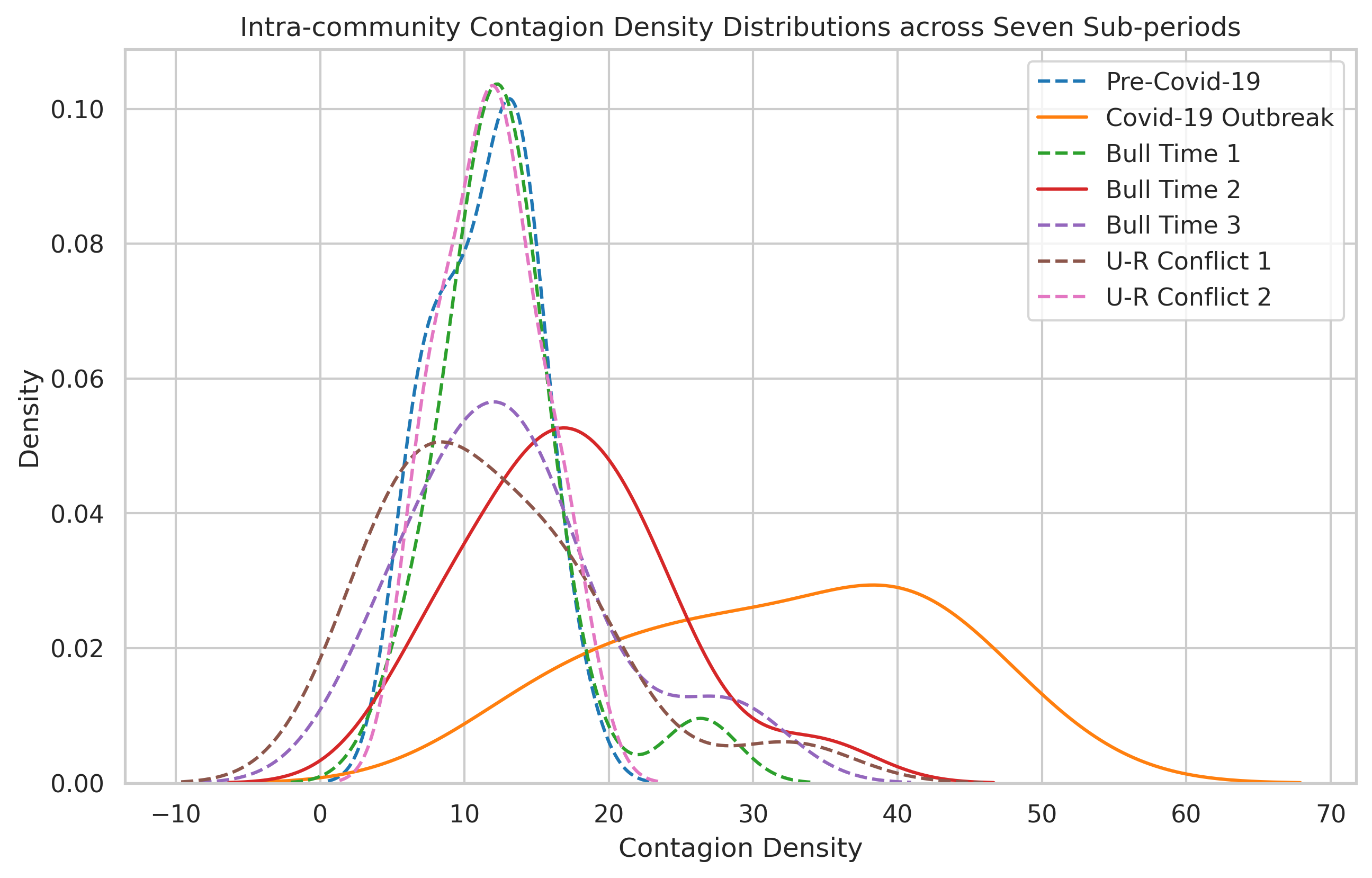}
\caption{Contagion Density}\label{fig10.1}
\end{subfigure}%
\begin{subfigure}{0.5\linewidth}
\centering
\includegraphics[width= 8.3cm]{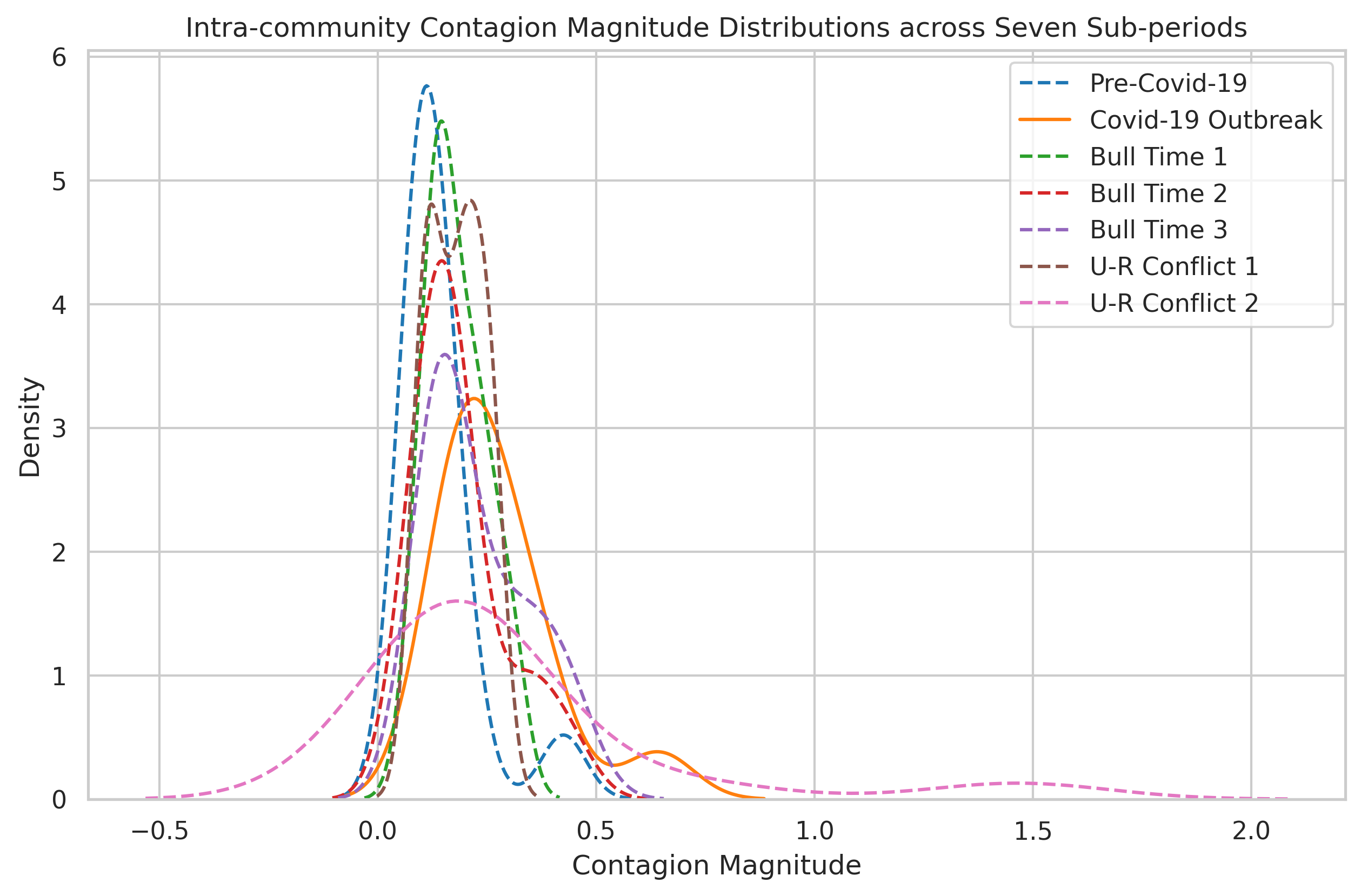}
\caption{Contagion Magnitude}\label{fig10.2}
\end{subfigure}

\caption{Distributions of  a) intra-community contagion densities and b) intra-community contagion magnitudes during each of seven sub-periods, namely Pre-Covid-19, Covid-19 Outbreak, Bull Time 1, Bull Time 2, Bull Time 3, Ukraine-Russia 1 and Ukraine-Russia 2. We use solid lines to highlight the distributions of notable sub-periods. Otherwise, they are drawn by dashed lines \label{fig:10}}
\end{figure}
\begin{figure}[h!]
\centering
\begin{subfigure}{0.5\linewidth}
\centering
\includegraphics[width= 8.5cm]{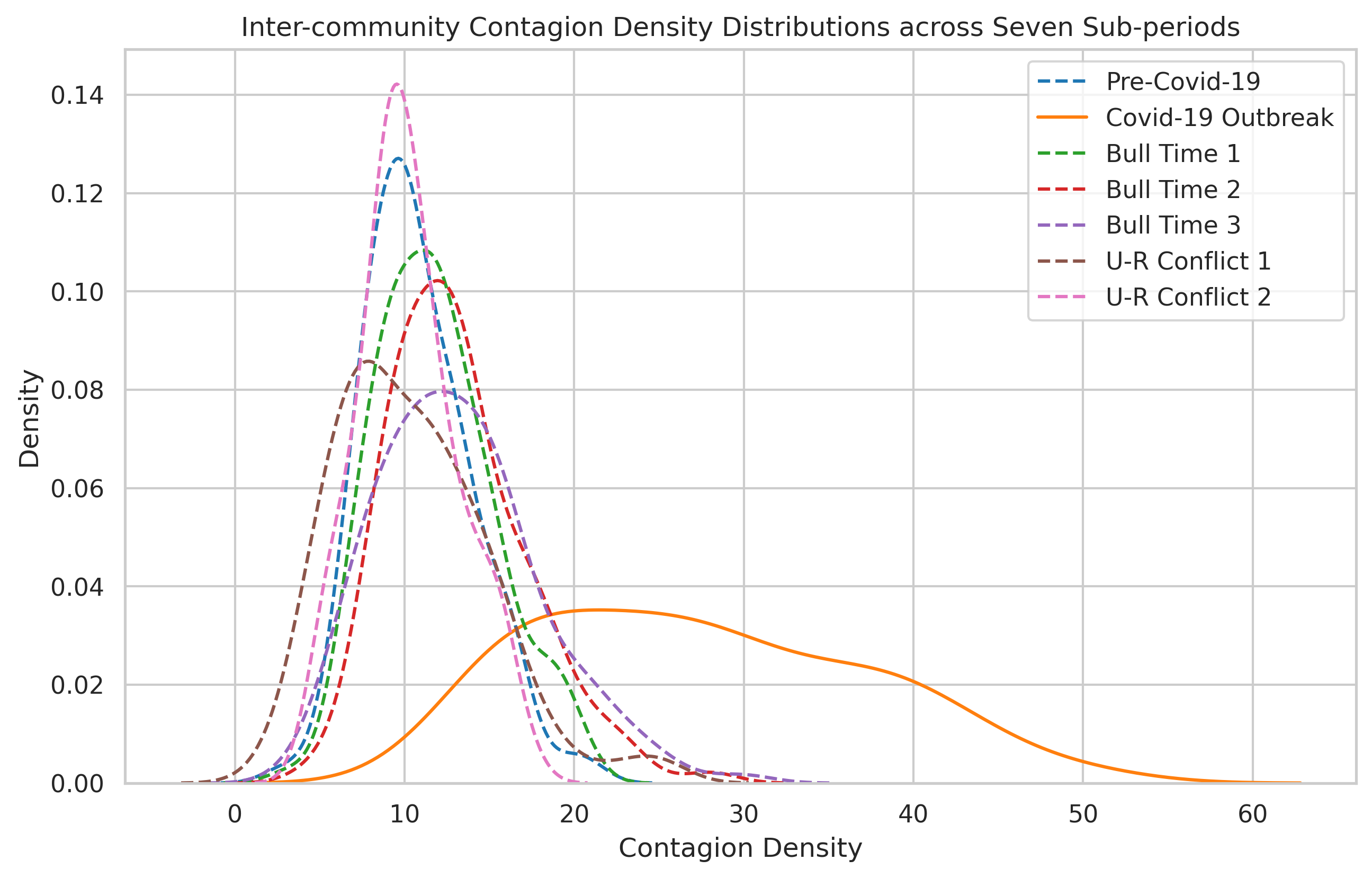}
\caption{Contagion Density}\label{fig11.1}
\end{subfigure}%
\begin{subfigure}{0.5\linewidth}
\centering
\includegraphics[width= 8.3cm]{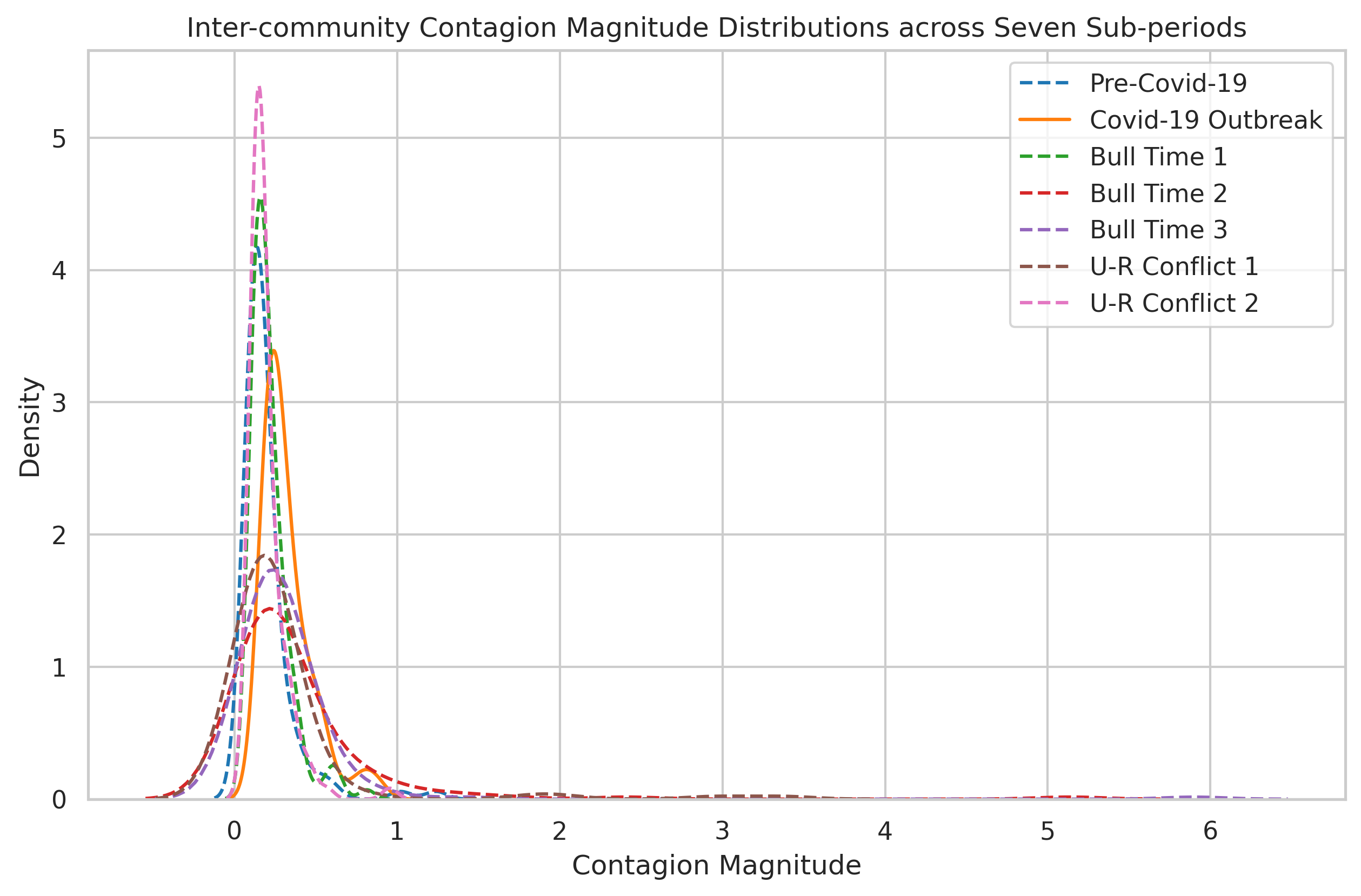}
\caption{Contagion Magnitude}\label{fig11.2}
\end{subfigure}

\caption{Distributions of  a) inter-community contagion densities and b) inter-community contagion magnitudes during each of seven sub-periods, namely Pre-Covid-19, Covid-19 Outbreak, Bull Time 1, Bull Time 2, Bull Time 3, Ukraine-Russia 1 and Ukraine-Russia 2. We use solid lines to highlight the distributions of notable sub-periods, otherwise, they are drawn by dashed lines \label{fig:11}}
\end{figure}

We first discuss the changes in contagion effects across the examined sub-periods. Specifically, Figures \ref{fig:10} and \ref{fig:11} display the distributions of intra-community and inter-community contagion metrics (e.g. density (a) and magnitude (b)) across seven sub-periods, respectively. Each sub-period is drawn in a different color. Moreover, we purposely highlight the distributions of some sub-periods that behave relatively differently from the rest using solid lines to help readers better locate them in the figures, while other sub-periods are marked by dashed lines. We find that both intra-community and inter-community contagion densities are dramatically amplified during the Covid-19 Outbreak period (solid orange line), compared to other time intervals, as shown in Figures \ref{fig10.1} and \ref{fig11.1}. This corresponds to one of the most turbulent phases in financial markets, characterized by sharp price declines, heightened volatility, and widespread investor pessimism~\cite{wor22}. This result suggests that the amplified turbulence in financial markets and the extreme reactions of investors (e.g. purchasing and selling herding behavior) caused by the pandemic trigger contagion not only within communities of highly similar assets but also across different asset communities. Furthermore, the Bull Time 2 sub-period (solid red line) also shows a moderate increase in contagion density, although this phenomenon is much weaker than the Covid-19 Outbreak time. This was a period when most assets experienced a price surge, indicating that contagion is not only a feature of negative shocks but can also be triggered under favorable market conditions. In fact, the Covid-19 pandemic had a much stronger impact on financial markets than the bullish time, with widespread systemic disruptions and heightened investor attention~\cite{ngu25}, explaining the more pronounced contagion effect observed during that time.

To this end, a stronger contagion density during turbulent periods compared to tranquil times can be attributed to the interplay of two key factors. First, the heightened investor sensitivity during periods of market turbulence increases the investors' demand for timely updates in terms of financial markets. This, in turn, accelerates the dissemination of information among financial assets driven by social media, news, financial reports, and other channels \cite{yar22}. These conditions are more likely to lead to herding behavior coupled with lead-lag trading activities where investors may adjust their positions in one asset based on the historical price movements of another, thereby intensifying contagion effects throughout financial assets.~\cite{boh08,eco23,kin21}. 

Remarkably, the distributions of contagion magnitudes across all sub-periods remain relatively similar, indicating a consistent level of magnitude over time, regardless of market conditions. This pattern holds true for both intra-community and inter-community contagions, suggesting that while the number of contagion links tends to increase during turbulent times, the average strength of these links remains stable. This phenomenon seems to be attributed to the portfolio diversification and risk-hedging strategy from investors in which they adjust their holdings across multiple assets rather than only focus on a particular one, especially during a financial crisis \cite{sch00}. These widespread but diluted approaches can raise the scale of contagion without amplifying the strength of individual spillovers.

To reinforce this result, we introduce two contagion metrics based on the measures defined in Subsection \ref{sec:4.6}, including Total Contagion Density (TCD) and Total Contagion Magnitude (TCM). These metrics allow us to quantitatively measure and compare contagion density and magnitude across sub-periods. Specifically, for each sub-period, we compute the total contagion by summing the average intra-community and inter-community contagion, separately for density and magnitude. Details are as follows:
\begin{enumerate}
    \item Total Contagion Density (TCD): to measure the total average contagion density within a given period of time by considering both intra-community and inter-community contagion.\\
    Let $SD_{Intra}, SD_{Inter}$ be sets of intra-community and inter-community contagion densities within a period of time, respectively. Let $n_{SD}, m_{SD}$ be the cardinality of $SD_{Intra}, SD_{Inter}$, respectively.
    \begin{equation}
        TCD = \frac{\sum_{i \in SD_{Intra}}i}{n_{SD}} + \frac{\sum_{j \in SD_{Inter}}j}{m_{SD}} \label{eq:16}
    \end{equation}

    \item Total Contagion Magnitude (TCM): to measure the total average contagion magnitude within a given period of time by considering both intra-community and inter-community contagion. \\
    Let $SM_{Intra}, SM_{Inter}$ be sets of intra-community and inter-community contagion magnitudes within a period of time, respectively. Let $n_{SM}, m_{SM}$ be the cardinality of $SM_{Intra}, SM_{Inter}$, respectively.
    \begin{equation}
        TCM = \frac{\sum_{i \in SM_{Intra}}i}{n_{SM}} + \frac{\sum_{j \in SM_{Inter}}j}{m_{SM}} \label{eq:17}
    \end{equation}
    
\end{enumerate}

Table \ref{tab:revise_2} reports the total contagion density and magnitude for each sub-period. As can be seen, the TCD value during the Covid-19 Outbreak is substantially higher than in any other sub-periods, reaching 59.28, whereas most other sub-periods fall within the range of 22 to 26. Additionally, the results also indicate that the Bull Time 2 sub-period exhibits slightly elevated total contagion density compared with the remaining periods, although still far below the Covid-19 Outbreak level. Regarding the total contagion magnitude, although the Covid-19 Outbreak again records the highest TCM value, its difference from other sub-periods is relatively small since most of them show relatively similar levels. To this end, these findings support the patterns observed in Figures \ref{fig:11} and \ref{fig:12}: contagion density tends to amplify during turbulent periods while contagion magnitude appears much less sensitive to market conditions.

\begin{table}[h!]
\caption{Total contagion density and magnitude in each sub-period}
\label{tab:revise_2}
\centering
\resizebox{1.0\textwidth}{!}{
\begin{tabular}{cccccccc}
\hline
\textbf{Measure} & \textbf{Pre-Covid-19} & \textbf{\begin{tabular}[c]{@{}c@{}}Covid-19\\ Outbreak\end{tabular}} & \textbf{Bull Time 1} & \textbf{Bull Time 2} & \textbf{Bull Time 3} & \textbf{\begin{tabular}[c]{@{}c@{}}U-R\\ Conflict 1\end{tabular}} & \textbf{\begin{tabular}[c]{@{}c@{}}U-R\\ Conflict 2\end{tabular}} \\ \hline
Total Contagion Magnitude & 0.336 & 0.596 & 0.399 & 0.565 & 0.547 & 0.472 & 0.499 \\
Total Contagion Density & 22.236 & 59.278 & 24.493 & 30.339 & 26.577 & 22.414 & 22.079 \\ \hline
\end{tabular}}
\end{table}

We now compare the contagion effects between intra-community and inter-community during each sub-period. Utilizing the distribution of inter-community contagions described above (shown in Figure \ref{fig:11}), our idea for this experiment is to compare each intra-community contagion with the distribution to see if the intra-community one behaves similarly to the inter-community contagions or not. In this regard, an intra-community contagion is considered to be significantly stronger than inter-community contagions within a sub-period if its quantile value falls within the right-hand side rejection area in the inter-community contagion distribution, with the significance level being 1\% (i.e. the quantile value is greater than or equal to the 99th). Otherwise, it behaves similarly to inter-community contagions. For this, Tables \ref{tab:4}-\ref{tab:5} show the contagion density and magnitude, respectively, within each community along with their quantiles (inside the brackets) in the corresponding inter-community contagion distribution for each of the seven sub-periods. Generally, both tables reveal that the intra-community contagion effects do not look much different from inter-community contagion effects since most intra-community contagion signals are similar to inter-community contagion signals. Specifically, the density (Table \ref{tab:4}) and magnitude (Table \ref{tab:5}) values within each community mostly belong to the 99\% confidence interval of their corresponding inter-community contagion distribution and are outside the rejection areas, indicating that intra-community contagion tends to follow the same behavior as inter-community contagion, across different time intervals and market conditions. By contrast, only a few communities in certain sub-periods reveal an abnormal behavior where their intra-community contagion densities or magnitudes are significantly stronger than the others and are off the inter-community contagion distributions. This phenomenon is more frequent for contagion density.  For example, community \#9 shows a contagion density exceeding 26\% in Bull Time 1 and 32\% in Ukraine-Russia Conflict 1, placing it at the 100th percentile of the corresponding inter-community contagion density distributions for those periods. Similarly, communities \#7 (100th percentile) and \#11 (99th percentile) in Bull Time 2 also exhibit unusually high contagion densities. As for contagion magnitude, only communities \#1 and \#14 in the Ukraine-Russia Conflict 2 sub-period surpass inter-community levels. A common characteristic of these communities is that the assets within each community mostly come from the same business sector, either Information Technology, Financials or cryptocurrency.  For this result, a possible reason is that these assets are closely relevant to each other, sharing common fundamental drivers such as macroeconomic variables, regulatory changes and investors' behaviors. This similarity in risk exposures and underlying characteristics naturally leads to higher correlation in price movements, which tends to trigger contagion within the group, especially during market uncertainties~\cite{che22}. However, these communities only account for a very small part of the network. 

\begin{table}[h!]
\centering
\caption{Contagion density of each community and its corresponding quantile (inside a bracket) across seven sub-periods. The quantiles are based on the distribution of inter-community contagion densities during each sub-period. Density values with quantile exceeding or being equal to 0.99 are highlighted in bold.}
\label{tab:4}
\resizebox{0.8\textwidth}{!}{
\begin{tabular}{cccccccc}
\hline
\textbf{Community} & \textbf{Pre-Covid-19} & \textbf{\begin{tabular}[c]{@{}c@{}}Covid-19\\ Outbreak\end{tabular}} & \textbf{Bull Time 1} & \textbf{Bull Time 2} & \textbf{Bull Time 3} & \textbf{\begin{tabular}[c]{@{}c@{}}U-R\\ Conflict 1\end{tabular}} & \textbf{\begin{tabular}[c]{@{}c@{}}U-R\\ Conflict 2\end{tabular}} \\ \hline
1 & \begin{tabular}[c]{@{}c@{}}7.08\\ (0.086)\end{tabular} & \begin{tabular}[c]{@{}c@{}}39.50\\ (0.885)\end{tabular} & \begin{tabular}[c]{@{}c@{}}13.19\\ (0.683)\end{tabular} & \begin{tabular}[c]{@{}c@{}}21.21\\ (0.958)\end{tabular} & \begin{tabular}[c]{@{}c@{}} \textbf{28.79}\\ \textbf{(0.993)}\end{tabular} & \begin{tabular}[c]{@{}c@{}}5.56\\ (0.133)\end{tabular} & \begin{tabular}[c]{@{}c@{}}7.60\\ (0.171)\end{tabular} \\ 
2 & \begin{tabular}[c]{@{}c@{}}14.71\\ (0.867)\end{tabular} & \begin{tabular}[c]{@{}c@{}}44.17\\ (0.974)\end{tabular} & \begin{tabular}[c]{@{}c@{}}5.77\\ (0.008)\end{tabular} & \begin{tabular}[c]{@{}c@{}}5.88\\ (0.008)\end{tabular} & \begin{tabular}[c]{@{}c@{}}15.93\\ (0.765)\end{tabular} & \begin{tabular}[c]{@{}c@{}}10.46\\ (0.514)\end{tabular} & \begin{tabular}[c]{@{}c@{}}11.04\\ (0.648)\end{tabular} \\ 
3 & \begin{tabular}[c]{@{}c@{}}10.29\\ (0.505)\end{tabular} & \begin{tabular}[c]{@{}c@{}}32.75\\ (0.692)\end{tabular} & \begin{tabular}[c]{@{}c@{}}11.11\\ (0.467)\end{tabular} & \begin{tabular}[c]{@{}c@{}}17.90\\ (0.867)\end{tabular} & \begin{tabular}[c]{@{}c@{}}11.36\\ (0.390)\end{tabular} & \begin{tabular}[c]{@{}c@{}}16.67\\ (0.924)\end{tabular} & \begin{tabular}[c]{@{}c@{}}12.50\\ (0.800)\end{tabular} \\ 
4 & \begin{tabular}[c]{@{}c@{}}13.33\\ (0.810)\end{tabular} & \begin{tabular}[c]{@{}c@{}}40.95\\ (0.910)\end{tabular} & \begin{tabular}[c]{@{}c@{}}12.09\\ (0.575)\end{tabular} & \begin{tabular}[c]{@{}c@{}}15.56\\ (0.750)\end{tabular} & \begin{tabular}[c]{@{}c@{}}10.42\\ (0.301)\end{tabular} & \begin{tabular}[c]{@{}c@{}}9.09\\ (0.457)\end{tabular} & \begin{tabular}[c]{@{}c@{}}7.14\\ (0.133)\end{tabular} \\ 
5 & \begin{tabular}[c]{@{}c@{}}6.67\\ (0.057)\end{tabular} & \begin{tabular}[c]{@{}c@{}}38.57\\ (0.846)\end{tabular} & \begin{tabular}[c]{@{}c@{}}13.64\\ (0.733)\end{tabular} & \begin{tabular}[c]{@{}c@{}}9.17\\ (0.150)\end{tabular} & \begin{tabular}[c]{@{}c@{}}8.89\\ (0.199)\end{tabular} & \begin{tabular}[c]{@{}c@{}}7.14\\ (0.229)\end{tabular} & \begin{tabular}[c]{@{}c@{}}7.14\\ (0.133)\end{tabular} \\ 
6 & \begin{tabular}[c]{@{}c@{}}6.36\\ (0.057)\end{tabular} & \begin{tabular}[c]{@{}c@{}}19.21\\ (0.269)\end{tabular} & \begin{tabular}[c]{@{}c@{}}7.50\\ (0.083)\end{tabular} & \begin{tabular}[c]{@{}c@{}}15.26\\ (0.742)\end{tabular} & \begin{tabular}[c]{@{}c@{}}15.46\\ (0.743)\end{tabular} & \begin{tabular}[c]{@{}c@{}}20.00\\ (0.971)\end{tabular} & \begin{tabular}[c]{@{}c@{}}12.08\\ (0.771)\end{tabular} \\ 
7 & \begin{tabular}[c]{@{}c@{}}12.88\\ (0.743)\end{tabular} & \begin{tabular}[c]{@{}c@{}}39.29\\ (0.872)\end{tabular} & \begin{tabular}[c]{@{}c@{}}13.19\\ (0.683)\end{tabular} & \begin{tabular}[c]{@{}c@{}} \textbf{34.44}\\ \textbf{(1.000)}\end{tabular} & \begin{tabular}[c]{@{}c@{}}\textbf{28.03}\\ \textbf{(0.993)}\end{tabular} & \begin{tabular}[c]{@{}c@{}}3.79\\ (0.019)\end{tabular} & \begin{tabular}[c]{@{}c@{}}10.00\\ (0.533)\end{tabular} \\ 
8 & \begin{tabular}[c]{@{}c@{}}13.68\\ (0.838)\end{tabular} & \begin{tabular}[c]{@{}c@{}}27.62\\ (0.538)\end{tabular} & \begin{tabular}[c]{@{}c@{}}12.22\\ (0.600)\end{tabular} & \begin{tabular}[c]{@{}c@{}}21.43\\ (0.958)\end{tabular} & \begin{tabular}[c]{@{}c@{}}5.56\\ (0.022)\end{tabular} & \begin{tabular}[c]{@{}c@{}}8.18\\ (0.381)\end{tabular} & \begin{tabular}[c]{@{}c@{}}16.07\\ (0.971)\end{tabular} \\ 
9 & \begin{tabular}[c]{@{}c@{}}17.31\\ (0.971)\end{tabular} & \begin{tabular}[c]{@{}c@{}}26.61\\ (0.500)\end{tabular} & \begin{tabular}[c]{@{}c@{}}\textbf{26.36}\\ \textbf{(1.000)}\end{tabular} & \begin{tabular}[c]{@{}c@{}}10.91\\ (0.342)\end{tabular} & \begin{tabular}[c]{@{}c@{}}13.19\\ (0.537)\end{tabular} & \begin{tabular}[c]{@{}c@{}}\textbf{32.50}\\ \textbf{(1.000)}\end{tabular} & \begin{tabular}[c]{@{}c@{}}16.19\\ (0.981)\end{tabular} \\ 
10 & \begin{tabular}[c]{@{}c@{}}8.18\\ (0.229)\end{tabular} & \begin{tabular}[c]{@{}c@{}}17.53\\ (0.205)\end{tabular} & \begin{tabular}[c]{@{}c@{}}9.85\\ (0.333)\end{tabular} & \begin{tabular}[c]{@{}c@{}}15.39\\ (0.742)\end{tabular} & \begin{tabular}[c]{@{}c@{}}14.39\\ (0.632)\end{tabular} & \begin{tabular}[c]{@{}c@{}}17.14\\ (0.933)\end{tabular} & \begin{tabular}[c]{@{}c@{}}12.08\\ (0.771)\end{tabular} \\ 
11 & \begin{tabular}[c]{@{}c@{}}14.84\\ (0.867)\end{tabular} & \begin{tabular}[c]{@{}c@{}}26.67\\ (0.526)\end{tabular} & \begin{tabular}[c]{@{}c@{}}14.55\\ (0.783)\end{tabular} & \begin{tabular}[c]{@{}c@{}}\textbf{25.46}\\ \textbf{(0.992)}\end{tabular} & \begin{tabular}[c]{@{}c@{}}10.26\\ (0.287)\end{tabular} & \begin{tabular}[c]{@{}c@{}}10.44\\ (0.514)\end{tabular} & \begin{tabular}[c]{@{}c@{}}\textbf{17.84}\\ \textbf{(1.000)}\end{tabular} \\ 
12 & \begin{tabular}[c]{@{}c@{}}12.88\\ (0.743)\end{tabular} & \begin{tabular}[c]{@{}c@{}}13.33\\ (0.026)\end{tabular} & \begin{tabular}[c]{@{}c@{}}10.00\\ (0.358)\end{tabular} & \begin{tabular}[c]{@{}c@{}}20.51\\ (0.950)\end{tabular} & \begin{tabular}[c]{@{}c@{}}13.33\\ (0.559)\end{tabular} & \begin{tabular}[c]{@{}c@{}}15.71\\ (0.905)\end{tabular} & \begin{tabular}[c]{@{}c@{}}14.29\\ (0.895)\end{tabular} \\
13 & \begin{tabular}[c]{@{}c@{}}9.62\\ (0.381)\end{tabular} & \begin{tabular}[c]{@{}c@{}}48.21\\ (0.987)\end{tabular} & \begin{tabular}[c]{@{}c@{}}10.95\\ (0.442)\end{tabular} & \begin{tabular}[c]{@{}c@{}}14.76\\ (0.725)\end{tabular} & \begin{tabular}[c]{@{}c@{}}15.46\\ (0.743)\end{tabular} & \begin{tabular}[c]{@{}c@{}}4.55\\ (0.048)\end{tabular} & \begin{tabular}[c]{@{}c@{}}12.09\\ (0.771)\end{tabular} \\ 
14 & \begin{tabular}[c]{@{}c@{}}10.44\\ (0.514)\end{tabular} &  & \begin{tabular}[c]{@{}c@{}}12.28\\ (0.608)\end{tabular} & \begin{tabular}[c]{@{}c@{}}16.36\\ (0.808)\end{tabular} & \begin{tabular}[c]{@{}c@{}}3.79\\ (0.007)\end{tabular} & \begin{tabular}[c]{@{}c@{}}14.76\\ (0.838)\end{tabular} & \begin{tabular}[c]{@{}c@{}}12.73\\ (0.800)\end{tabular} \\ 
15 & \begin{tabular}[c]{@{}c@{}}13.07\\ (0.762)\end{tabular} &  & \begin{tabular}[c]{@{}c@{}}16.99\\ (0.917)\end{tabular} & \begin{tabular}[c]{@{}c@{}}22.53\\ (0.975)\end{tabular} & \begin{tabular}[c]{@{}c@{}}8.82\\ (0.191)\end{tabular} & \begin{tabular}[c]{@{}c@{}}4.44\\ (0.048)\end{tabular} & \begin{tabular}[c]{@{}c@{}}8.93\\ (0.343)\end{tabular} \\ 
16 &  &  & \begin{tabular}[c]{@{}c@{}}13.74\\ (0.733)\end{tabular} & \begin{tabular}[c]{@{}c@{}}9.09\\ (0.150)\end{tabular} & \begin{tabular}[c]{@{}c@{}}20.51\\ (0.919)\end{tabular} &  &  \\ 
17 &  &  &  &  & \begin{tabular}[c]{@{}c@{}}4.44\\ (0.007)\end{tabular} &  &  \\ \hline
\end{tabular}}
\end{table}

\begin{table}[h!]
\centering
\caption{Contagion magnitude of each community and its corresponding quantile (inside a bracket) across seven sub-periods. The quantiles are based on the distribution of inter-community contagion magnitudes during each sub-period. Magnitude values with quantile exceeding or being equal to 0.99 are highlighted in bold.}
\label{tab:5}
\resizebox{0.8\textwidth}{!}{
\begin{tabular}{cccccccc}
\hline
\textbf{Community} & \textbf{Pre-Covid-19} & \textbf{\begin{tabular}[c]{@{}c@{}}Covid-19\\ Outbreak\end{tabular}} & \textbf{Bull Time 1} & \textbf{Bull Time 2} & \textbf{Bull Time 3} & \textbf{\begin{tabular}[c]{@{}c@{}}U-R\\ Conflict 1\end{tabular}} & \textbf{\begin{tabular}[c]{@{}c@{}}U-R\\ Conflict 2\end{tabular}} \\ \hline
1 & \begin{tabular}[c]{@{}c@{}}0.13\\ (0.410)\end{tabular} & \begin{tabular}[c]{@{}c@{}}0.39\\ (0.744)\end{tabular} & \begin{tabular}[c]{@{}c@{}}0.30\\ (0.833)\end{tabular} & \begin{tabular}[c]{@{}c@{}}0.22\\ (0.483)\end{tabular} & \begin{tabular}[c]{@{}c@{}}0.33\\ (0.772)\end{tabular} & \begin{tabular}[c]{@{}c@{}}0.26\\ (0.800)\end{tabular} & \begin{tabular}[c]{@{}c@{}} \textbf{0.70}\\ \textbf{(0.990)}\end{tabular} \\ 
2 & \begin{tabular}[c]{@{}c@{}}0.12\\ (0.333)\end{tabular} & \begin{tabular}[c]{@{}c@{}}0.24\\ (0.333)\end{tabular} & \begin{tabular}[c]{@{}c@{}}0.10\\ (0.025)\end{tabular} & \begin{tabular}[c]{@{}c@{}}0.07\\ (0.000)\end{tabular} & \begin{tabular}[c]{@{}c@{}}0.14\\ (0.103)\end{tabular} & \begin{tabular}[c]{@{}c@{}}0.11\\ (0.029)\end{tabular} & \begin{tabular}[c]{@{}c@{}}0.09\\ (0.010)\end{tabular} \\ 
3 & \begin{tabular}[c]{@{}c@{}}0.43\\ (0.943)\end{tabular} & \begin{tabular}[c]{@{}c@{}}0.25\\ (0.410)\end{tabular} & \begin{tabular}[c]{@{}c@{}}0.15\\ (0.308)\end{tabular} & \begin{tabular}[c]{@{}c@{}}0.14\\ (0.200)\end{tabular} & \begin{tabular}[c]{@{}c@{}}0.40\\ (0.853)\end{tabular} & \begin{tabular}[c]{@{}c@{}}0.26\\ (0.781)\end{tabular} & \begin{tabular}[c]{@{}c@{}}0.09\\ (0.000)\end{tabular} \\ 
4 & \begin{tabular}[c]{@{}c@{}}0.06\\ (0.000)\end{tabular} & \begin{tabular}[c]{@{}c@{}}0.64\\ (0.949)\end{tabular} & \begin{tabular}[c]{@{}c@{}}0.30\\ (0.842)\end{tabular} & \begin{tabular}[c]{@{}c@{}}0.08\\ (0.000)\end{tabular} & \begin{tabular}[c]{@{}c@{}}0.15\\ (0.140)\end{tabular} & \begin{tabular}[c]{@{}c@{}}0.19\\ (0.571)\end{tabular} & \begin{tabular}[c]{@{}c@{}}0.18\\ (0.610)\end{tabular} \\ 
5 & \begin{tabular}[c]{@{}c@{}}0.16\\ (0.552)\end{tabular} & \begin{tabular}[c]{@{}c@{}}0.21\\ (0.231)\end{tabular} & \begin{tabular}[c]{@{}c@{}}0.14\\ (0.258)\end{tabular} & \begin{tabular}[c]{@{}c@{}}0.10\\ (0.042)\end{tabular} & \begin{tabular}[c]{@{}c@{}}0.46\\ (0.890)\end{tabular} & \begin{tabular}[c]{@{}c@{}}0.25\\ (0.752)\end{tabular} & \begin{tabular}[c]{@{}c@{}}0.15\\ (0.400)\end{tabular} \\ 
6 & \begin{tabular}[c]{@{}c@{}}0.09\\ (0.057)\end{tabular} & \begin{tabular}[c]{@{}c@{}}0.35\\ (0.705)\end{tabular} & \begin{tabular}[c]{@{}c@{}}0.08\\ (0.000)\end{tabular} & \begin{tabular}[c]{@{}c@{}}0.36\\ (0.742)\end{tabular} & \begin{tabular}[c]{@{}c@{}}0.21\\ (0.441)\end{tabular} & \begin{tabular}[c]{@{}c@{}}0.13\\ (0.152)\end{tabular} & \begin{tabular}[c]{@{}c@{}}0.30\\ (0.857)\end{tabular} \\ 
7 & \begin{tabular}[c]{@{}c@{}}0.12\\ (0.333)\end{tabular} & \begin{tabular}[c]{@{}c@{}}0.34\\ (0.679)\end{tabular} & \begin{tabular}[c]{@{}c@{}}0.15\\ (0.308)\end{tabular} & \begin{tabular}[c]{@{}c@{}}0.13\\ (0.175)\end{tabular} & \begin{tabular}[c]{@{}c@{}}0.31\\ (0.735)\end{tabular} & \begin{tabular}[c]{@{}c@{}}0.12\\ (0.057)\end{tabular} & \begin{tabular}[c]{@{}c@{}}0.27\\ (0.819)\end{tabular} \\ 
8 & \begin{tabular}[c]{@{}c@{}}0.18\\ (0.648)\end{tabular} & \begin{tabular}[c]{@{}c@{}}0.18\\ (0.064)\end{tabular} & \begin{tabular}[c]{@{}c@{}}0.23\\ (0.700)\end{tabular} & \begin{tabular}[c]{@{}c@{}}0.16\\ (0.342)\end{tabular} & \begin{tabular}[c]{@{}c@{}}0.15\\ (0.140)\end{tabular} & \begin{tabular}[c]{@{}c@{}}0.11\\ (0.029)\end{tabular} & \begin{tabular}[c]{@{}c@{}}0.15\\ (0.400)\end{tabular} \\ 
9 & \begin{tabular}[c]{@{}c@{}}0.11\\ (0.229)\end{tabular} & \begin{tabular}[c]{@{}c@{}}0.27\\ (0.487)\end{tabular} & \begin{tabular}[c]{@{}c@{}}0.13\\ (0.208)\end{tabular} & \begin{tabular}[c]{@{}c@{}}0.14\\ (0.217)\end{tabular} & \begin{tabular}[c]{@{}c@{}}0.12\\ (0.037)\end{tabular} & \begin{tabular}[c]{@{}c@{}}0.19\\ (0.571)\end{tabular} & \begin{tabular}[c]{@{}c@{}}0.20\\ (0.733)\end{tabular} \\ 
10 & \begin{tabular}[c]{@{}c@{}}0.13\\ (0.352)\end{tabular} & \begin{tabular}[c]{@{}c@{}}0.17\\ (0.038)\end{tabular} & \begin{tabular}[c]{@{}c@{}}0.16\\ (0.367)\end{tabular} & \begin{tabular}[c]{@{}c@{}}0.15\\ (0.250)\end{tabular} & \begin{tabular}[c]{@{}c@{}}0.29\\ (0.684)\end{tabular} & \begin{tabular}[c]{@{}c@{}}0.21\\ (0.657)\end{tabular} & \begin{tabular}[c]{@{}c@{}}0.21\\ (0.743)\end{tabular} \\ 
11 & \begin{tabular}[c]{@{}c@{}}0.07\\ (0.000)\end{tabular} & \begin{tabular}[c]{@{}c@{}}0.23\\ (0.282)\end{tabular} & \begin{tabular}[c]{@{}c@{}}0.15\\ (0.308)\end{tabular} & \begin{tabular}[c]{@{}c@{}}0.18\\ (0.375)\end{tabular} & \begin{tabular}[c]{@{}c@{}}0.10\\ (0.015)\end{tabular} & \begin{tabular}[c]{@{}c@{}}0.18\\ (0.543)\end{tabular} & \begin{tabular}[c]{@{}c@{}}0.26\\ (0.819)\end{tabular} \\ 
12 & \begin{tabular}[c]{@{}c@{}}0.08\\ (0.038)\end{tabular} & \begin{tabular}[c]{@{}c@{}}0.15\\ (0.000)\end{tabular} & \begin{tabular}[c]{@{}c@{}}0.22\\ (0.642)\end{tabular} & \begin{tabular}[c]{@{}c@{}}0.19\\ (0.408)\end{tabular} & \begin{tabular}[c]{@{}c@{}}0.15\\ (0.147)\end{tabular} & \begin{tabular}[c]{@{}c@{}}0.20\\ (0.638)\end{tabular} & \begin{tabular}[c]{@{}c@{}}0.12\\ (0.190)\end{tabular} \\ 
13 & \begin{tabular}[c]{@{}c@{}}0.20\\ (0.714)\end{tabular} & \begin{tabular}[c]{@{}c@{}}0.15\\ (0.000)\end{tabular} & \begin{tabular}[c]{@{}c@{}}0.19\\ (0.575)\end{tabular} & \begin{tabular}[c]{@{}c@{}}0.33\\ (0.708)\end{tabular} & \begin{tabular}[c]{@{}c@{}}0.13\\ (0.044)\end{tabular} & \begin{tabular}[c]{@{}c@{}}0.24\\ (0.743)\end{tabular} & \begin{tabular}[c]{@{}c@{}}0.22\\ (0.752)\end{tabular} \\ 
14 & \begin{tabular}[c]{@{}c@{}}0.11\\ (0.267)\end{tabular} &  & \begin{tabular}[c]{@{}c@{}}0.13\\ (0.200)\end{tabular} & \begin{tabular}[c]{@{}c@{}}0.20\\ (0.417)\end{tabular} & \begin{tabular}[c]{@{}c@{}}0.13\\ (0.051)\end{tabular} & \begin{tabular}[c]{@{}c@{}}0.12\\ (0.048)\end{tabular} & \begin{tabular}[c]{@{}c@{}} \textbf{1.47}\\ \textbf{(1.000)}\end{tabular} \\ 
15 & \begin{tabular}[c]{@{}c@{}}0.09\\ (0.057)\end{tabular} &  & \begin{tabular}[c]{@{}c@{}}0.22\\ (0.633)\end{tabular} & \begin{tabular}[c]{@{}c@{}}0.17\\ (0.117)\end{tabular} & \begin{tabular}[c]{@{}c@{}}0.19\\ (0.353)\end{tabular} & \begin{tabular}[c]{@{}c@{}}0.10\\ (0.010)\end{tabular} & \begin{tabular}[c]{@{}c@{}}0.12\\ (0.190)\end{tabular} \\ 
16 &  &  & \begin{tabular}[c]{@{}c@{}}0.25\\ (0.733)\end{tabular} & \begin{tabular}[c]{@{}c@{}}0.44\\ (0.800)\end{tabular} & \begin{tabular}[c]{@{}c@{}}0.40\\ (0.853)\end{tabular} &  &  \\ 
17 &  &  &  &  & \begin{tabular}[c]{@{}c@{}}0.21\\ (0.441)\end{tabular} &  &  \\ \hline
\end{tabular}}
\end{table}

To further assess the similarity between intra-community and inter-community contagion effects, we apply the Kolmogorov–Smirnov test \cite{mas51}, which evaluates whether two samples are drawn from the same underlying distribution. A statistically significant result indicates that the two distributions differ, whereas a p-value greater than the chosen significance level suggests that the samples are statistically indistinguishable and therefore similar in their characteristics. In this present study, we set the confidence level to be 99\%, which is equivalent to a significance level of 0.01. P-values of the Kolmogorov-Smirnov tests between intra-community and inter-community contagion effects during different sub-periods are shown in Table \ref{tab:r2.3}.

\begin{table}[h!]
\caption{P-values of the Kolmogorov–Smirnov tests comparing intra-community and inter-community contagion in each sub-period. A p-value below 1\% (0.01) indicates that the two distributions differ, whereas a p-value above this threshold suggests no statistical difference between them. Contagion is evaluated in two aspects: density and magnitude. A star symbol denotes tests with p-values below 0.01.}
\label{tab:r2.3}
\resizebox{\textwidth}{!}{
\begin{tabular}{cccccccc}
\hline
\textbf{\begin{tabular}[c]{@{}c@{}}Contagion \\ Measure\end{tabular}} & \textbf{Pre-Covid-19} & \textbf{\begin{tabular}[c]{@{}c@{}}Covid-19\\ Outbreak\end{tabular}} & \textbf{Bull Time 1} & \textbf{Bull Time 2} & \textbf{Bull Time 3} & \textbf{\begin{tabular}[c]{@{}c@{}}U-R\\ Conflict 1\end{tabular}} & \textbf{\begin{tabular}[c]{@{}c@{}}U-R\\ Conflict 2\end{tabular}} \\ \hline
Density & 0.272 & 0.210 & 0.513 & \begin{tabular}[c]{@{}c@{}}0.002*\\ 0.023 (adjusted)\end{tabular} & 0.826 & 0.402 & 0.052 \\
Magnitude & 0.107 & 0.475 & 0.589 & 0.069 & 0.070 & 0.236 & 0.272 \\ \hline
\end{tabular}}
\end{table}

Across all sub-periods, almost all p-values exceed 0.01, indicating no statistically significant difference between intra-community and inter-community contagion in both density and magnitude. The only exception occurs in the Bull Time 2 sub-period, where the contagion density between intra-community and inter-community shows a statistically significant difference, with p-values below 0.01. A closer examination reveals that this result is driven by only a few communities exhibiting relatively high contagion density, as mentioned and explained earlier. Once these communities are removed from the analysis, the p-value rises above the significance threshold, confirming the similarity between intra-community and inter-community contagion. Thus, even during this sub-period, the majority of communities display contagion patterns consistent with those observed across communities.

Another noteworthy finding is that although cryptocurrencies are highly correlated with each other and exhibit distinct behavior from other traditional assets, as shown in Subsection \ref{sec:5.1}, the contagion density and magnitude within communities of cryptocurrencies generally do not differ from other asset groups. The only notable exception is during the Bull Time 2 sub-period, when the contagion density among cryptocurrencies stood at the highest level across the entire network, exceeding both intra- and inter-community contagion densities. As discussed earlier, cryptocurrencies share several common fundamental drivers, especially the strong sensitivity of investors to positive market conditions \cite{ngu25}, combined with the positive momentum of financial markets during this bullish time, which explains this high contagion density.

Overall, the above results suggest that financial contagion tends to manifest similarly both within and across communities since there is no significant difference between intra-community and inter-community contagions. This implies that contagion is unlikely to be an important factor in the formation of communities. Supporting this point is the case of cryptocurrencies: although they exhibit strong internal correlations and form distinct communities, their contagion levels are generally comparable to those of other asset communities and cross-communities throughout most periods.

\subsection{Directional Inter-community Contagion} \label{sec:5.3}
In this subsection, we take a closer look at inter-community contagion effects by analyzing how shocks propagate directionally between different communities. Specifically, we investigate how one community transmits shocks to another and how it, in turn, receives shocks. Our goal is to identify whether there exist communities acting as dominant transmitters (e.g. showing strong outgoing contagion to a high number of communities) or receivers (showing strong incoming contagion from a high number of communities). Also, as discussed in the previous subsection, contagion magnitude shows rather small variation over time and offers few meaningful insights. Therefore, for the sake of clarity and simplicity, we focus solely on contagion density in this part of the analysis. To this end, we use the directional contagion density metric described in Equation \ref{eq:14} for this experiment.

\begin{figure}[h!]
\centering
\begin{subfigure}{0.51\linewidth}
\centering
\includegraphics[width= 7cm]{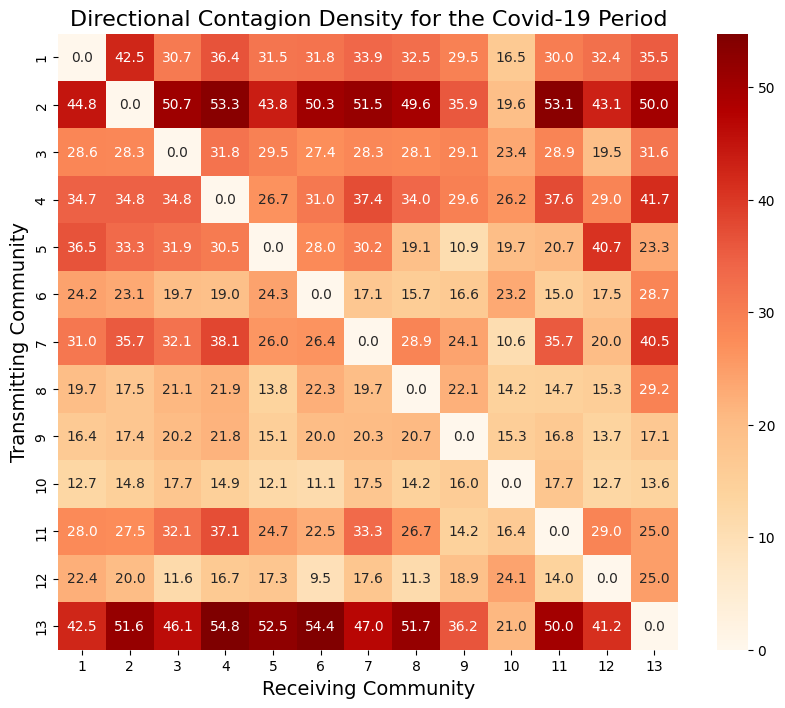}
\caption{Contagion Heatmap}\label{fig12.1}
\end{subfigure}%
\begin{subfigure}{0.5\linewidth}
\centering
\includegraphics[width= 7cm]{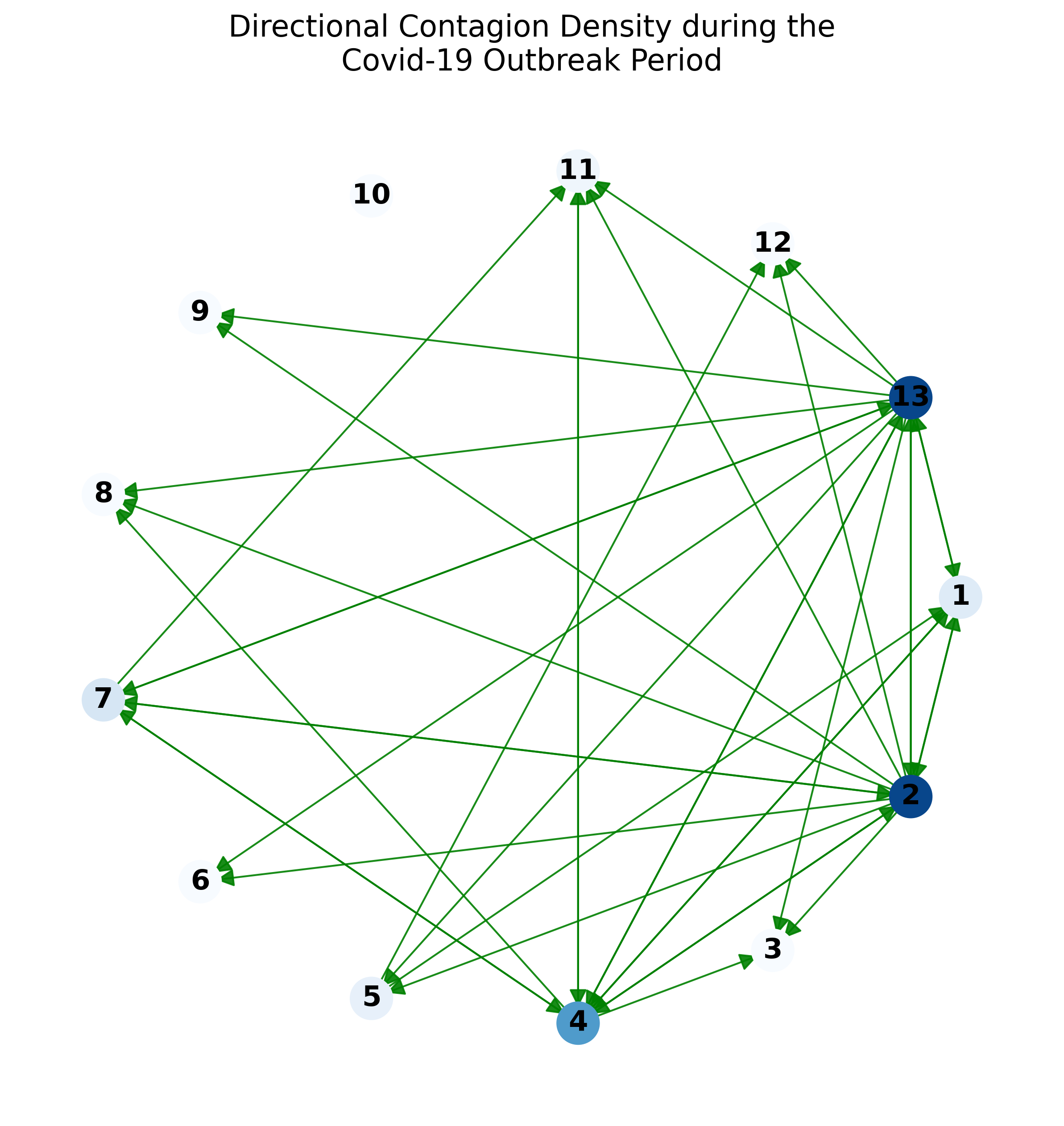}
\caption{Contagion Graph}\label{fig12.2}
\end{subfigure}%
\caption{Heatmap of directional inter-community contagion density (a) and its corresponding graph representation of top 25\% in contagion density (b). In both sub-figures, a darker color indicates a higher value. An arrow represents the direction of contagion from one community to another. \label{fig:12}}
\end{figure}

To visualize the directional contagion density between communities, we employ heatmaps. For this, each row of the heatmap represents a transmitting community while each column represents a receiving community. For example, the cell at  position [1,2] reflects the contagion effect from community \#1 to community \#2. The intensity of the color corresponds to the strength of the contagion density - the darker the color, the higher the density. We note that only inter-community contagions are considered, so the diagonal values are set to zero. At first sight, a noticeable pattern emerging across all sub-periods is that certain communities act as dominant transmitters in the network, dispersing shocks to multiple other communities at significantly higher densities than the rest. An example of this phenomenon is displayed in Figure \ref{fig12.1}, corresponding to the Covid-19 Outbreak sub-period. Specifically, communities \#2 and \#13 exhibit high contagion density, transmitting shocks to nearly all remaining communities, as emphasized by their darker red rows in the heatmap compared to the rest. To further investigate this behavior, we construct a directed graph for each sub-period based on the corresponding heatmap, retaining only the top 25\% of contagion links in terms of density. This filtering highlights the most prominent contagion channels within the network. In this graph, each node represents a community while an edge between two nodes exists only if the contagion density between them ranks in the top 25\%. Each edge is directional, indicated by an arrow pointing from the transmitting to the receiving community. Moreover, each node is colored based on the number of outgoing contagion links, i.e. the higher the number of outgoing contagion links, the darker the color. An example of this graph is shown in Figure \ref{fig12.2}, which is equivalent to the directional contagion density heatmap during the Covid-19 Outbreak sub-period displayed in Figure \ref{fig12.1}. Additionally, quantitative details about these graph representations are also calculated where each graph is summarized in tabular form, reporting the number of top-25\% incoming and outgoing contagion links for each community, the average contagion density of incoming contagion links to a community and the  average contagion density of outgoing contagion links from a community. The graphs and their corresponding quantitative details across all sub-periods are reported in the supporting file S4.

\begin{figure}[h!]
\centering
\begin{subfigure}{0.5\linewidth}
\centering
\includegraphics[width= 7cm]{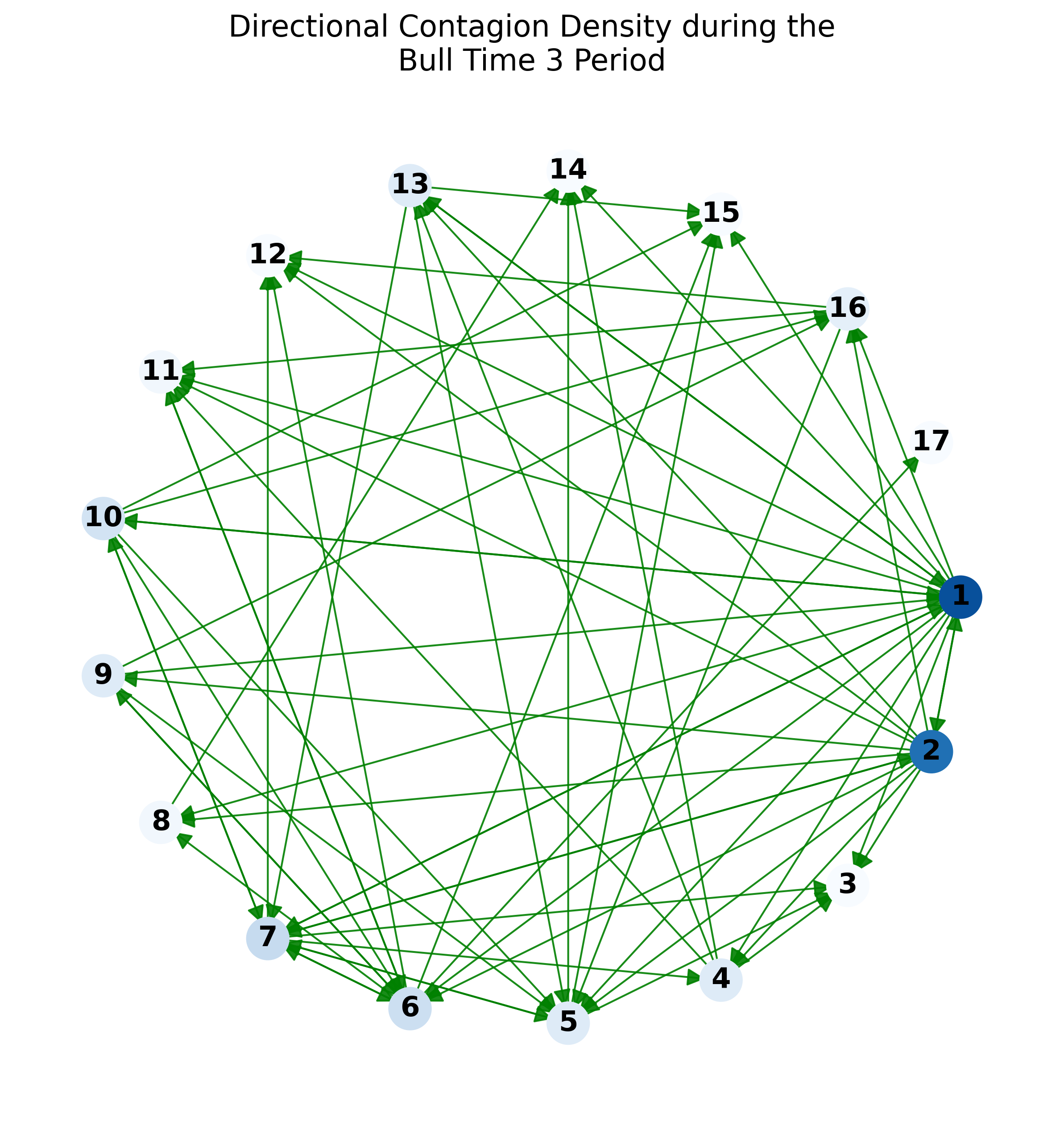}
\caption{Bull Time 3}\label{fig13.1}
\end{subfigure}%
\begin{subfigure}{0.5\linewidth}
\centering
\includegraphics[width= 7cm]{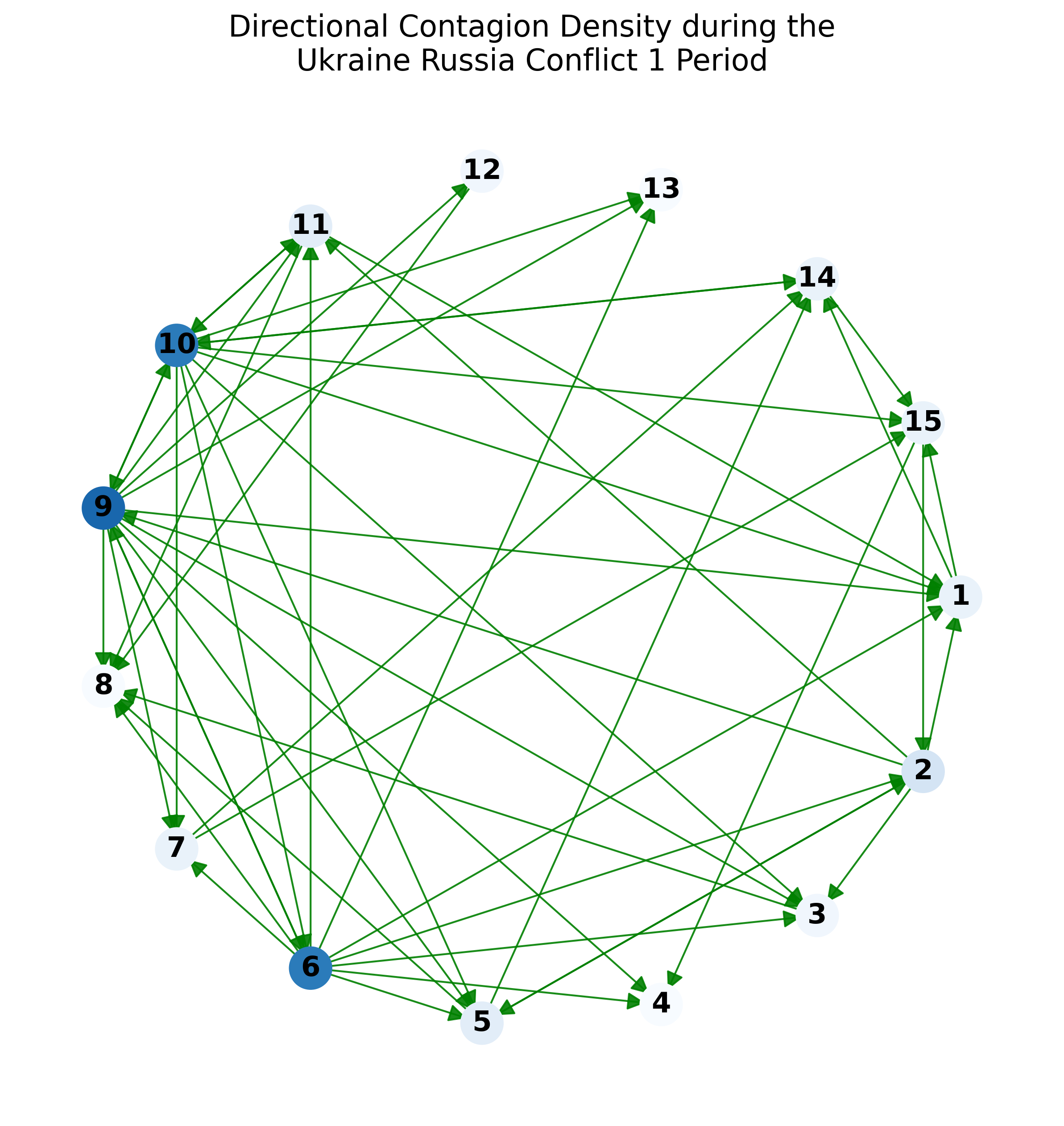}
\caption{Ukraine-Russia Conflict 1}\label{fig13.2}
\end{subfigure}%

\caption{Graph representation of the top 25\% inter-community contagions in terms of the contagion density during the Bull Time 3 (a) and Ukraine-Russia Conflict 1 (b) sub-periods. A darker color indicates a higher value. An arrow represents the direction of contagion from one community to another. \label{fig:13}}
\end{figure}

Indeed, our experimental results show that there are always communities acting as strong transmitters in each sub-period, spreading shocks to many other communities with significantly higher contagion density levels relative to the overall network. On the other hand, this pattern for the case of receiving communities is not observed across all sub-periods, suggesting that there is no community intensively receiving shocks from many others. Notably, communities that act as strong transmitters in the network are often composed of stocks and US ETFs dominated by holdings in the Information Technology sector. Besides, sectors like Healthcare, Financials, Energy and cryptocurrencies also emerge as strong transmitters in specific periods. To support these findings, we present two examples: the graph visualizations of the top 25\% inter-community contagions in terms of contagion density (Figures \ref{fig13.1} and \ref{fig13.2}) and their quantitative details (Tables \ref{tab:6.1} and \ref{tab:6.2}) for the Bull Time 3 and Ukraine-Russia Conflict 1 sub-periods, respectively. Specifically, Table \ref{tab:6.1} shows that communities \#1 and \#2 during the Bull Time 3 sub-period act as dominant transmitters, transmitting high-density shocks (i.e. belong to the top 25\% of inter-community contagions in terms of the contagion density) to 14 and 12 out of 16 communities, respectively. These numbers are significantly higher than those for other communities. Moreover, the average contagion densities from communities \#1 and \#2  to other communities also stand at the highest levels, with 23.75\% and 20.10\%, respectively. Consequently, these two communities are highlighted in the darkest blue in Figure \ref{fig13.1}. Notably, community \#1 is composed predominantly of stocks and US ETFs dominated by holdings in Information Technology (75\%) while Healthcare assets contribute approximately 86\% to community \#2. Likewise, during the Ukraine-Russia Conflict 1 sub-period, communities \#6, \#9 and \#10 emerge as dominant transmitters. For this, Financials assets primarily make up community \#9, whereas Information Technology assets account for the majority in the other two communities. Further details are provided in Table \ref{tab:6.2}. Beyond these examples, a summary of the dominant transmitters in each sub-period and their primary business sectors is provided in the supporting file S4.

\begin{table}[h!]
\centering
\caption{Quantitative details of the top 25\% inter-community contagion links in terms of density during the (a) Bull Time 3 and (b) Ukraine-Russia Conflict 1 sub-periods. \textit{Com} represents the community, \textit{\#In} is the number of incoming contagion links to a community, \textit{\#Out} is the number of outgoing contagion links from a community, \textit{Avg\_In} is the average contagion density of incoming contagion links to a community, and \textit{Avg\_Out} is the average contagion density of outgoing contagion links from a community. Communities acting as dominant transmitters are highlighted in bold.}\label{tab:6}
  \begin{minipage}{0.5\linewidth}
    \centering
\resizebox{0.95\textwidth}{!}{%
\begin{tabular}{ccccc}
\hline
\textbf{Com} & \textbf{\#In} & \textbf{\#Out} & \textbf{Avg\_In (\%)} & \textbf{Avg\_Out (\%)} \\ \hline
\textbf{1} & \textbf{5} & \textbf{14} & \textbf{14.40} & \textbf{23.75}  \\ 
\textbf{2} & \textbf{2} & \textbf{12} & \textbf{12.57} & \textbf{20.10}  \\ 
3 & 5 & 0 & 12.87 & 9.11  \\ 
4 & 3 & 4 &  12.96 & 13.99 \\ 
5 & 7 & 4  & 15.87 & 12.51 \\ 
6 & 6 & 7 & 14.01  & 16.50 \\
7 & 6 & 8  & 14.63 & 16.73 \\ 
8 & 3 & 1  & 11.83 & 11.64 \\ 
9 & 2 & 4  & 9.81 & 14.53 \\ 
10 & 2 & 6  & 11.45 & 15.00 \\ 
11 & 5 & 1 & 13.07 & 12.19  \\ 
12 & 5 & 0  & 14.51 & 7.13 \\ 
13 & 3 & 4 & 13.91  & 14.06 \\ 
14 & 4 & 0  & 14.66 & 6.11 \\ 
15 & 5 & 0  & 13.48 & 7.95 \\ 
16 & 4 & 3  & 12.87 & 14.61 \\ 
17 & 1 & 0  & 10.25 & 7.26 \\ \hline
\end{tabular} 
}%
    \subcaption{Bull Time 3}\label{tab:6.1}
  \end{minipage}%
  \begin{minipage}{0.5\linewidth}
    \centering
    \resizebox{0.95\textwidth}{!}{%
\begin{tabular}{ccccc}
\hline
\textbf{Com} & \textbf{\#In} & \textbf{\#Out} & \textbf{Avg\_In (\%)} & \textbf{Avg\_Out (\%)} \\ \hline
1 & 5 & 2  & 10.76 & 8.54 \\ 
2 & 3 & 5  & 9.45 & 11.35 \\ 
3 & 4 & 1 & 11.75  & 10.06 \\ 
4 & 3 & 0  & 11.39 & 7.49 \\ 
5 & 4 & 3  & 10.71 & 8.66 \\ 
\textbf{6} & \textbf{2} & \textbf{10}  & \textbf{9.22} & \textbf{17.44} \\ 
7 & 3 & 2  & 10.48 & 4.95  \\ 
8 & 6 & 0  & 11.50 & 8.34 \\ 
\textbf{9} & \textbf{3} & \textbf{11}  & \textbf{10.63} & \textbf{20.80} \\ 
\textbf{10} & \textbf{3} & \textbf{10}  & \textbf{10.52} & \textbf{17.78} \\ 
11 & 4 & 3  & 10.79 & 9.61 \\ 
12 & 1 & 1  & 7.28 & 8.97 \\ 
13 & 3 & 0  & 9.53 & 6.71 \\ 
14 & 4 & 2  & 10.82 & 6.95 \\ 
15 & 4 & 2 & 10.94  & 8.13 \\ \hline
\end{tabular}
}%
\vspace{0.8cm}
    \subcaption{Ukraine-Russia Conflict 1} \label{tab:6.2}
  \end{minipage}
  
\end{table}

The strong influence of these sectors can be attributed to their intrinsic characteristics and also the prevailing market conditions during the examined periods. In particular, Information Technology assets consistently attract significant investor attention due to their strong growth potential and dominant market capitalization in the financial markets, making them highly influential in shaping both investor behavior and overall market performance. As a result, price movements in this sector often have a broad impact across various business sectors and types of investment vehicle. In fact, the sector’s influence has grown even more remarkably since 2020 thanks to the accelerated global technology adoption followed by the boom in generative AI~\cite{flo24,rud23}. Regarding other aforementioned sectors, their strong influence appears to be more context-dependent. For instance,  during the pandemic, the Healthcare sector gained prominence as companies were at the forefront of efforts to combat Covid-19, including the production of protective equipment and vaccines~\cite{bow22,ami23}. This rapid development elevated their importance and impact on both financial markets and the broader economy. Likewise, the Financials sector plays a vital role in reflecting and shaping economic conditions~\cite{mar23}, thereby carrying a strong influence in financial markets, especially during periods of uncertainty like the Ukraine-Russia conflict. To this end, our results are supported to some extent by existing studies, although they are not directly relevant to our present topic and dataset context. For instance, Giudici and Parisi in \cite{giu18} examined the systemic risk among ten countries using Corporate Default Swap (CDS) spreads over a period from 2006 to 2016. By introducing a novel approach that integrates partial correlations and correlation networks into VAR models, they discovered a clustering effect between the countries in terms of credit risk contagion. Specifically, they found that some countries tend to act as risk importers, while some  often behave as risk exporters. Despite the differences in the two topics, this finding shares some common ground with the results of our experiments.

Overall, when combining these results with those from the previous subsection, they suggest that while contagion might not be an important factor causing the formation of communities, the community structure in general and some communities in particular might still be significantly influenced by certain communities through their significant contagion effects, highlighting their central role in both the contagion network and the financial markets as a whole. Although this hypothesis requires further validation through broader experiments, the current findings offer a promising first step in exploring this topic. 

\section{Robustness Analyses}\label{sec:6} 
In this section, we run robustness analyses of our results when using other approaches in conducting experiments. We cover a wide range of experiments to prove that our results are consistent even when different approaches in different parts of our experiments are selected. These include: 1) evaluating the community structure (i.e. asset partition) using an alternative random matrix theory to filter noise signals; 2) reassessing contagion results under a stricter 99\% confidence level; 3) examining the community structure derived from an alternative network construction—the Planar Maximally Filtered Graph (PMFG).

\subsection{Marchenko-Pastur Random Matrix Theory for Noise Filtering} \label{sec:6.1}

Our noise filtering procedure in this study relies on the Tracy-Widom random matrix theory. Since the purpose of this preprocessing step is to find the asset partition (or community structure) without the effect of noise in financial markets,  we evaluate the consistency of the resulting partition with that obtained using the Marchenko–Pastur random matrix theory \cite{ngu22,giu22}. In this regard, we adopt the v-measure metric \cite{ngu22} to assess the similarity between two community structures obtained by the two aforementioned methods. This metric ranges from 0 to 1, where a value of 1 indicates identical community structures while a value of 0 indicates completely dissimilar ones. Results for this comparison are shown in Table \ref{tab:r1.1}. Additionally, we also investigate the number of eigenvalues considered to be informative under each method to determine whether their selections differ. This result is displayed in Table \ref{tab:r1.2}.

\begin{table}[h!]
\caption{Community structure comparison between Tracy-Widom-based noise filtering and Marchenko-Pastur-based noise filtering.}
\label{tab:r1.1}
\resizebox{\textwidth}{!}{
\begin{tabular}{cccccccc}
\hline
\textbf{Measure} & \textbf{Pre-Covid-19} & \textbf{\begin{tabular}[c]{@{}c@{}}Covid-19\\ Outbreak\end{tabular}} & \textbf{Bull Time 1} & \textbf{Bull Time 2} & \textbf{Bull Time 3} & \textbf{\begin{tabular}[c]{@{}c@{}}U-R\\ Conflict 1\end{tabular}} & \textbf{\begin{tabular}[c]{@{}c@{}}U-R\\ Conflict 2\end{tabular}} \\ \hline
v-measure & 0.90 & 1.00 & 0.94 & 0.82 & 0.87 & 0.90 & 0.82 \\ \hline
\end{tabular}}
\end{table}

\begin{table}[h!]
\caption{Number of informative eigenvalues in the correlation matrix for each sub-period, as determined by the Tracy–Widom and Marchenko–Pastur methods.}
\centering
\label{tab:r1.2}
\begin{tabular}{ccc}
\hline
\textbf{Sub-period} & \textbf{Tracy-Widom} & \textbf{Marchenko-Pastur} \\ \hline
Pre-Covid-19 & 18 & 15 \\
Covid-19 Outbreak & 8 & 8 \\
Bull Time 1 & 15 & 13 \\
Bull Time 2 & 17 & 14 \\
Bull Time 3 & 16 & 13 \\
U-R Conflict 1 & 13 & 11 \\
U-R Conflict 2 & 16 & 13 \\ \hline
\end{tabular}
\end{table}

As shown in Table \ref{tab:r1.1}, the community structures obtained between two theories exhibit a high degree of similarity, with v-measure values exceeding 80\% in all sub-periods and most above 90\%. This indicates that the use of either Tracy-Widom or Marchenko-Pastur random matrix theory does not cause significant changes in the noise filtering result. However, looking at Table \ref{tab:r1.2}, we notice that the number of informative eigenvalues identified by Tracy-Widom theory tends to be higher than that obtained using the other. This occurs because Marchenko-Pastur relies on a strict theoretical formula to filter noise in the correlation matrix, making it sensitive to the sample size, as discussed earlier in Subsection \ref{sec:2.3}. As a result, some eigenvalues that carry meaningful information may still fall below the Marchenko-Pastur threshold due to their relatively small magnitude and are therefore classified as noise. In contrast, Tracy-Widom is less sensitive to data size and its threshold is based on empirical experimentation rather than a fixed theoretical distribution. This allows the theory to adapt more flexibly to the data and retain additional informative eigenvalues. 

It is worth noting that the Marchenko-Pastur random matrix theory has been widely adopted in previous studies because it preserves the most significant information while effectively filtering out noise. However, the more flexible and empirically driven nature of the Tracy-Widom one allows it to capture additional structure in the correlation matrix that Marchenko–Pastur may overlook by discarding small but informative eigenvalues.

\subsection{Contagion Coefficients Measured at the 99\% Confidence Level} \label{sec:6.2}

We are currently employing a 90\% confidence level to filter out unreliable contagion coefficients estimated from the VAR model, as discussed in Subsection \ref{sec:4.5}. Recall that each contagion coefficient in this study measures the extent to which movements in one asset influence another. Therefore, coefficients equal to 0.0 or non-zero coefficients that fail to meet the 90\% confidence threshold indicate no contagion relationship between the corresponding pair of assets. To test the robustness of the original contagion coefficients (i.e. at the 90\% confidence level), we evaluate them using a stricter confidence level of 99\%. Table \ref{tab:r2.2} reports the percentage of original non-zero contagion coefficients that are statistically significant at the 90\% level and still remain significant at the 99\% level.

\begin{table}[h!]
\caption{Percentage of the original non-zero contagion coefficients used in this study that are still statistically significant at the 99\% confidence level.}
\label{tab:r2.2}
\resizebox{\textwidth}{!}{
\begin{tabular}{ccccccc}
\hline
\textbf{Pre-Covid-19} & \textbf{\begin{tabular}[c]{@{}c@{}}Covid-19\\ Outbreak\end{tabular}} & \textbf{Bull Time 1} & \textbf{Bull Time 2} & \textbf{Bull Time 3} & \textbf{\begin{tabular}[c]{@{}c@{}}U-R\\ Conflict 1\end{tabular}} & \textbf{\begin{tabular}[c]{@{}c@{}}U-R\\ Conflict 2\end{tabular}} \\ \hline
90.07 & 94.96 & 91.17 & 91.29 & 91.38 & 90.84 & 90.97 \\ \hline
\end{tabular}}
\end{table}

Across all sub-periods, more than 90\% of the contagion coefficients that are statistically significant at the 90\% confidence level remain significant when evaluated at the stricter 99\% threshold. This implies that, in each sub-period, at most 10\% of the original coefficients are filtered out under the higher confidence level. Indeed, since each sub-period is sufficiently long for reliable VAR estimation, the resulting contagion coefficients generally have very low p-values, generally close to 0.0. Consequently, increasing the confidence level has a low impact on our original contagion coefficients.

To assess the robustness of our original findings under a stricter confidence threshold, we repeat the key experiments using the new 99\% confidence level and compare the new results with our original ones. These include: 1) examining the contagion effects within and between communities across all sub-periods to assess the evolution of contagion over time; 2) comparing intra-community and inter-community contagion; 3) analyzing directional contagions between communities. Overall, the main findings remain unchanged. However, contagion densities both within and between communities tend to be slightly lower than before. This decline is expected since raising the confidence level reduces the number of reliable contagion coefficients (i.e. the contagion links between assets). By contrast, the new contagion magnitudes do not exhibit this reduction. Details of these experiments are provided in the supporting file S5.

\subsection{Planar Maximally Filtered Graph (PMFG) for Network Construction} \label{sec:6.3}

\begin{figure}[h!]
\centering
\begin{subfigure}{\linewidth}
\centering
\includegraphics[width= 11cm]{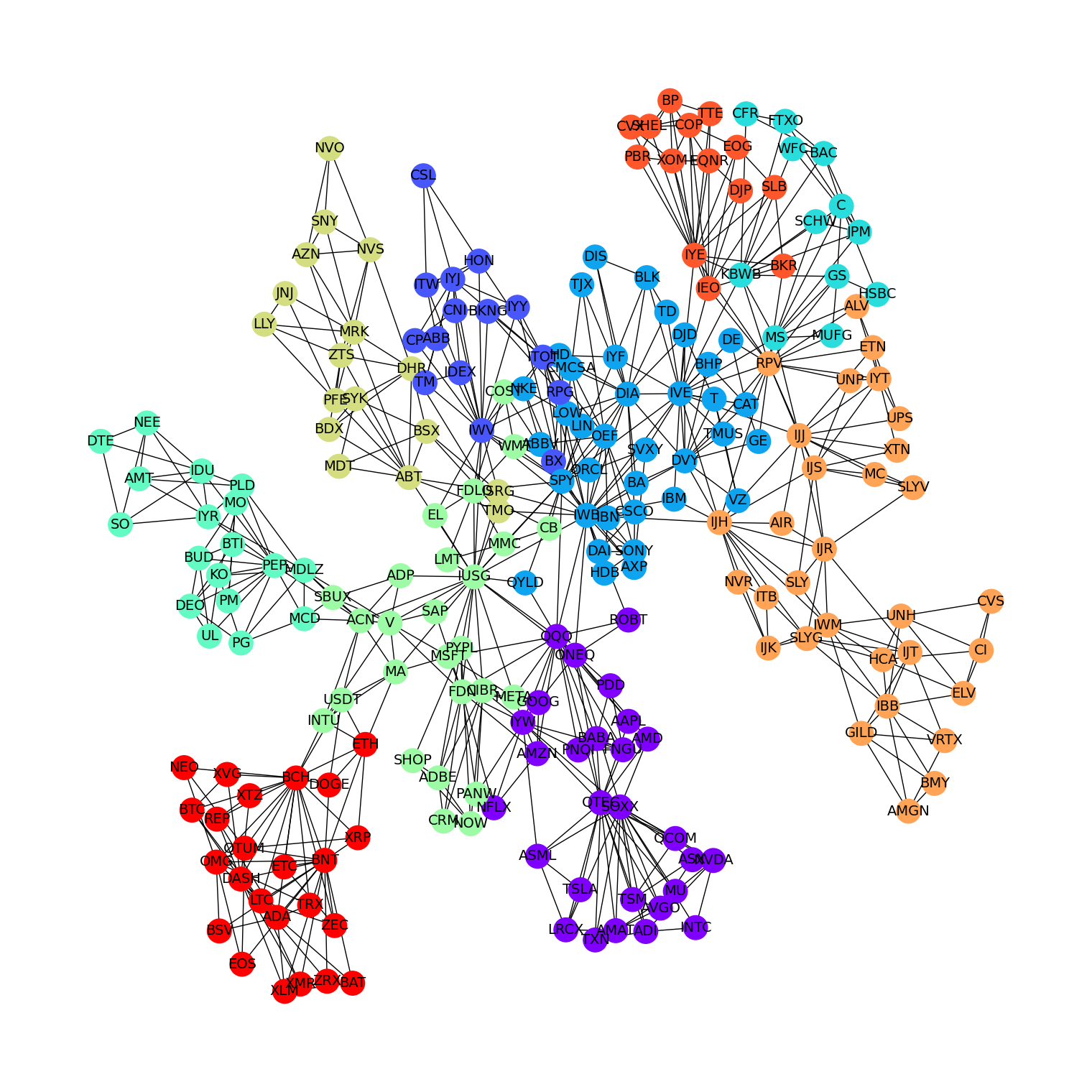}
\caption{Network and Community Structure}\label{fig9.1}
\end{subfigure}%

\begin{subfigure}{\linewidth}
\centering
\includegraphics[width= 11cm]{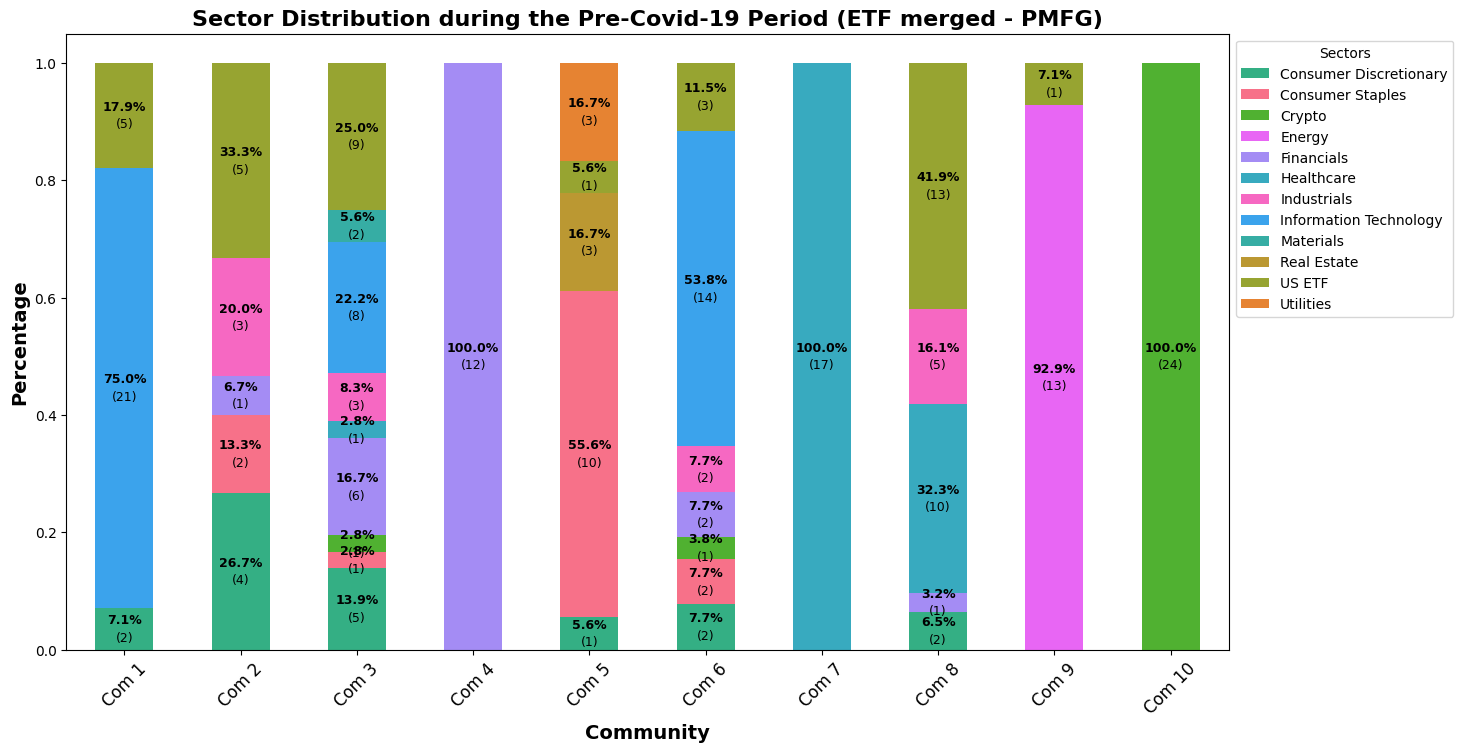}
\caption{Sector Distribution}\label{fig9.2}
\end{subfigure}

\caption{Network and Community structure (a) and sector distribution per community (b) of 221 assets during the Pre-Covid-19 period, based on the Planar Maximally Filtered Graph (PMFG). A US ETF is assigned to a specific business sector if it is designed to track the performance of stocks within that sector. Otherwise, it is unclassified and labeled as US ETF (i.e. US ETFs that do not track a specific sector or those that include stocks from multiple sectors. \label{fig:9}}
\end{figure}

In our study, we first construct an asset network using the Minimum Spanning Tree (MST) and then apply the Louvain community detection algorithm to identify its community structure. However, because the MST retains only N-1 edges, it may discard many potentially important relationships between financial assets. This reduction can bias the resulting communities and, in turn, lead to unreliable community-based experimental results. To address this concern, we examine the robustness of the MST-derived community structure  and its relevant contagion findings by comparing them with those obtained from another well-established network representation: the Planar Maximally Filtered Graph (PMFG) \cite{tum05}. In brief, this graph preserves significantly more information by maintaining planarity - meaning no edges intersect - while still retaining the strongest asset relationships. As a result, it offers a richer and more informative network with 3(N-2) edges. Figure \ref{fig9.1} presents the PMFG-based network and its corresponding community structure for the Pre–Covid-19 sub-period as an example. The sector distribution of each identified community is then displayed in Figure \ref{fig9.2}.

We begin by comparing the similarity between community structures derived from our original MST-based approach and those obtained using the PMFG, using the v-measure as the evaluation metric. Results are presented in Table \ref{tab:r2.1}, showing that the PMFG-based communities are largely consistent with those from the MST. Specifically, across all sub-periods and market conditions, the similarity remains high - averaging around 80\%, indicating that the community structure is relatively stable regardless of the underlying network construction method. Notably, we observe that the number of communities identified using the PMFG is consistently lower than that from the MST. This phenomenon arises from the denser network structure produced by the PMFG, which includes more connections between assets. As a result, each community tends to contain more assets. Table \ref{tab:r2.1.2} shows the number of communities in each sub-period, as detected from the MST and PMFG, respectively.

\begin{table}[h!]
\caption{Community structure comparison between MST and PMFG}
\label{tab:r2.1}
\resizebox{\textwidth}{!}{
\begin{tabular}{cccccccc}
\hline
\textbf{Measure} & \textbf{Pre-Covid-19} & \textbf{\begin{tabular}[c]{@{}c@{}}Covid-19\\ Outbreak\end{tabular}} & \textbf{Bull Time 1} & \textbf{Bull Time 2} & \textbf{Bull Time 3} & \textbf{\begin{tabular}[c]{@{}c@{}}U-R\\ Conflict 1\end{tabular}} & \textbf{\begin{tabular}[c]{@{}c@{}}U-R\\ Conflict 2\end{tabular}} \\ \hline
v-measure & 0.85 & 0.76 & 0.79 & 0.79 & 0.78 & 0.84 & 0.85 \\ \hline
\end{tabular}}
\end{table}

\begin{table}[h!]
\caption{Number of communities in each sub-period, as detected from MST and PMFG.}
\centering
\label{tab:r2.1.2}
\begin{tabular}{ccc}
\hline
\textbf{Sub-period} & \textbf{MST-based} & \textbf{PMFG-based} \\ \hline
Pre-Covid-19 & 15 & 10 \\
Covid-19 Outbreak & 13 & 8 \\
Bull Time 1 & 16 & 9 \\
Bull Time 2 & 16 & 10 \\
Bull Time 3 & 17 & 9 \\
U-R Conflict 1 & 15 & 9 \\
U-R Conflict 2 & 15 & 11 \\ \hline
\end{tabular}
\end{table}

To validate our original findings, we rerun the main contagion analyses similarly to the previous Subsection \ref{sec:6.2} using the community structure derived from the PMFG. Results from all experiments remain consistent with those obtained using the MST-based structure, confirming the robustness of our conclusions. To this end, although the PMFG yields a slightly different community structure as reflected in the differing number of communities, the main findings reported in this study are fully preserved. Detailed results of these experiments are provided in the supporting file S6.

\newpage
\section{Conclusion} \label{sec:7}

In this study, we discover contagion effects in a broad financial market comprising stocks, cryptocurrencies and US ETFs during a period when a number of impactful events were taking place. Our work contributes to the existing literature in different ways. First and foremost, the contagion is examined at the community (e.g. a group of assets) level rather than focusing solely on individual asset relationships. Next, we segment the initial full period in question into distinct sub-periods based on major financial events, allowing for a detailed exploration of contagion dynamics under varying market conditions. Moreover, unlike many prior studies that concentrate on a single type of investment vehicle, our analysis spans multiple types, offering a more comprehensive and cross-sectional view of financial contagion.

Observing intra- and inter-community contagion under various market conditions, including the turbulent time caused by the Covid-19 pandemic, bullish phases and periods of political conflict, we find that contagion density both within a community and between two communities tends to intensify during times of market uncertainty, as might be expected. This holds true regardless of whether the uncertainty stems from negative events such as the pandemic or positive developments such as strong market rallies. Notably, the degree of density appears to be proportional to the severity of market uncertainty. In contrast, the magnitude of contagion remains relatively stable across all market conditions. From a systemic risk perspective, this result indicates that market uncertainty is characterized more by the widespread diffusion of shocks than by the intensity of any single spillover. Therefore, one should pay attention to the expansion of contagion pathways, rather than focusing solely on impact magnitudes.

For the first research question, when comparing the intra-community with inter-community contagion, we find no significant difference in contagion density or magnitude between the two across all sub-periods. This suggests that contagion is a system-wide phenomenon rather than being constrained to specific communities or asset groups. Moreover, contagion is less likely to be a key factor in the formation of communities. Consequently, adopting portfolio diversification by investing in assets from different communities or assets with low correlations might offer limited protection against systemic contagion.

For the second research question, our analysis of directional contagion density across community pairs reveals that certain communities, particularly those composed of stocks and US ETFs dominated by holdings in the Information Technology sector, act as dominant contagion transmitters in the network, during each sub-period. This strong influence is likely driven by the sector’s strong growth prospects and substantial market capitalization. Additionally, communities forming mostly from Healthcare, Financials, Energy and cryptocurrency assets also emerge as dominant transmitters during specific sub-periods, possibly caused by prevailing market conditions. This result emphasizes the necessity for investors to monitor the Information Technology sector closely as its price movements or shocks may ripple through a wide range of assets. Furthermore, major events and their associated business sectors should be carefully monitored since they pose a heightened risk of widespread contagion throughout the market coming from the sectors.

While this study provides valuable insights into the financial contagion, several rooms for future research remain open. Firstly, the analysis could be extended to a broader market by including a larger set of assets and more types of investment vehicle. Secondly, expanding the analysis to include up-to-date data could provide additional insights to this study. Thirdly, incorporating exogenous variables such as macroeconomic indicators, monetary policy changes and sentiment scores from news and social media may help uncover the drivers behind contagion effects. Fourthly, another promising direction is to develop forecasting frameworks that incorporate contagion information to enhance forecasting accuracy. 

\section*{CRediT authorship contribution statement}
\textbf{An Pham Ngoc Nguyen}: Writing – review \& editing, Writing – original draft, Visualization, Validation, Software, Project administration, Methodology, Investigation, Formal analysis, Data curation, Conceptualization. \textbf{Marija Bezbradica}: Writing – review \& editing, Validation, Supervision, Resources, Project administration, Investigation, Funding acquisition, Conceptualization. \textbf{Martin Crane}: Writing – review \& editing, Validation, Supervision, Resources, Project administration, Investigation, Funding acquisition, Conceptualization.

\section*{Declaration of competing interest}
The authors declare that they have no known competing financial interests or personal relationships that could have appeared to influence the work reported in this paper.

\section*{Acknowledgments}
This research was conducted with the financial support of Taighde Éireann-Research Ireland under Grant Agreement No. 13/RC/2106\_P2 at the ADAPT Centre at Dublin City University. ADAPT, the Research Ireland Centre for AI-Driven Digital Content Technology, is funded through the Research Ireland Centres Programme. For the purpose of Open Access, the author has applied a CC BY public copyright licence to any Author Accepted Manuscript version arising from this submission.

\section*{Appendix A. Supplementary data}
Supplementary material related to this article can be found online at

\textcolor{blue}{https://doi.org/10.1016/j.chaos.2025.117858.}

\section*{Data availability}
Data is available in my Github repo: 

\textcolor{blue}{https://github.com/NguyenPhamNgocAn/Community-Level-Contagion-in-Financial-Markets.}

\scriptsize
\bibliographystyle{elsarticle-num}

\end{document}